\documentclass{aa}  

\usepackage{graphicx}
\usepackage{amsmath}   
\usepackage{amssymb}   
\usepackage{xcolor}
\usepackage{lscape}
\usepackage{longtable}
\usepackage{hyperref}
\usepackage{siunitx}
\usepackage[normalem]{ulem}

\usepackage{txfonts}
\usepackage{natbib}
\usepackage[version=4]{mhchem}
\bibpunct{(}{)}{;}{a}{}{,}
\begin{document}

 \title{A transition between the hot and the ultra-hot Jupiter atmospheres}

  \author{Claire Baxter \inst{1} \and
          Jean-Michel D\'esert \inst{1}\and
          Vivien Parmentier \inst{2}\and
          Mike Line \inst{3}\and
          Jonathan Fortney \inst{4}\and
          Jacob Arcangeli \inst{1}\and
          Jacob L. Bean \inst{5}\and
          Kamen O. Todorov \inst{1}\and
          Megan Mansfield \inst{6}
          }

  \institute{Anton Pannekoek Institute for Astronomy, University of Amsterdam, Science Park 904, 1098 XH Amsterdam, The Netherlands \\ \email{c.j.baxter@uva.nl} \and
              Department of Physics, University of Oxford, Oxford, OX1 3PU, UK \and
              School of Earth \& Space Exploration, Arizona State University, Tempe AZ 85287, USA \and
              Department of Astronomy and Astrophysics, University of California, Santa Cruz, CA 95064, USA \and
              Department of Astronomy \& Astrophysics, University of Chicago, 5640 S. Ellis Avenue, Chicago, IL 60637, USA \and 
              Department of Geophysical Sciences, University of Chicago, 5734 S. Ellis Ave., Chicago, IL 60637, USA
             }

   \date{Received Month day, Year; accepted Month day, Year}

 
  \abstract
  {A key hypothesis in the field of exoplanet atmospheres is the trend of atmospheric thermal structure with planetary equilibrium temperature. We explore this trend and report here the first statistical detection of a transition in the near-infrared (NIR) atmospheric emission between  hot and ultra-hot Jupiters. We measure this transition using secondary eclipse observations and interpret this phenomenon as changes in atmospheric properties, and more specifically in terms of transition from  non-inverted to inverted thermal profiles.
  We examine a sample of 78 hot Jupiters with secondary eclipse measurements at 3.6~$\mu$m and 4.5~$\mu$m measured with Spitzer Infrared Array Camera (IRAC). We calculate the planetary brightness temperatures using PHOENIX models to correct for the stellar flux. We measure the deviation of the data from the blackbody, which we define as the difference between the observed 4.5~$\mu$m eclipse depth and that expected at this wavelength based on the brightness temperature measured at 3.6~$\mu$m.   
  We study how the deviation between 3.6 and 4.5~$\mu$m changes with theoretical predictions with equilibrium temperature and incoming stellar irradiation.
  We reveal a clear transition in the observed emission spectra of the hot Jupiter population at $1660\pm100$~K in the zero albedo, full redistribution equilibrium temperature. We find the hotter exoplanets have even hotter daysides at 4.5~$\mu$m compared to 3.6~$\mu$m, which manifests as an exponential increase in the emitted power of the planets with stellar insolation.
  We propose that the measured transition is a result of seeing carbon monoxide in emission due to the formation of temperature inversions in the atmospheres of the hottest planets. These thermal inversions could be caused by the presence of atomic and molecular species with high opacities in the optical and/or the lack of cooling species.

  Our findings are in remarkable agreement with a new grid of 1D radiative and convective models varying metallicity, carbon to oxygen ratio (C/O), surface gravity, and stellar effective temperature. We find that the population of hot Jupiters statistically disfavors high C/O planets (C/O$\geq0.85$).}

   \maketitle

\section{Introduction}

Observing the infrared secondary eclipse of transiting tidally locked hot Jupiters allows us to measure their  dayside thermal flux \citep[e.g.,][]{Charbonneau2005, Deming2005, Cowan2011b,Cowan2011,Triaud2014b, Schwartz2015, Schwartz2017, Zhang2018,Garhart2020}. The dayside flux is determined by the temperature pressure (T-P) profile and the atmospheric opacities. In turn, the T-P profile is determined by the albedo, heat redistribution, and  atmospheric opacities.
Hot Jupiters have equilibrium temperatures around 1500K. But recently, a newer class of hot Jupiters has emerged, the ultra-hot Jupiters (UHJ). Ultra-hot Jupiters have equilibrium temperatures in excess of 2000K and receive irradiation 10-100 times the insolation of other hot Jupiters \citep[e.g., Figure 9.][]{Parmentier2018}. There is evidence that they exhibit different atmospheric properties from their cooler counterparts \citep[e.g.,][]{Bell2017, Arcangeli2018, Mansfield2018, Parmentier2018, Kreidberg2018}. Investigations by \citet{Hubeny2003, Fortney2006} and \citet{Fortney2008} suggest that temperature inversions could appear in hot Jupiter atmospheres at temperatures as low as 1700K resulting from a fundamental change in atmospheric opacity due to TiO and VO \citep{Gandhi2019}. Furthermore, \citet{Thorngren2019} suggest that the deep atmospheres of these planets are so hot that TiO and VO are able to stay in the gas phase at $\sim$1700-2000K rather than being cold-trapped into clouds at depth.

Previous studies have looked for signatures of physical processes (chemistry, thermal inversions, redistribution, albedo, stellar activity) in a large sample of atmospheres, specifically by looking at the thermal eclipse measurements \citep{Knutson2010, Cowan2011, Triaud2014b, Schwartz2015, Schwartz2017, Zhang2018, Garhart2020, Keating2019, Melville2020}. \citet{Triaud2014b} created color-magnitude diagrams of planets with available Spitzer/IRAC eclipses in all four bandpasses (3.6, 4.5, 5.8, and 8.0~$\mu$m). They found that hot Jupiters lie closer to brown dwarf (MLT) colors than they do to blackbodies, (i.e., they do not have featureless spectra in the infrared).

Additional studies have focused on breaking the degeneracy between albedo and redistribution efficiency. \citet{Cowan2011b} perform a statistical study on the energy budget of 24 hot Jupiters with secondary eclipses in at least one infrared waveband (>0.8~$\mu$m) and, when available, phase variation measurements. They found the sample as a whole could be represented with low Bond albedos. Additionally, in combination with \citet{Zhang2018}, there is evidence of low redistribution efficiencies of the eight hottest planets (WASP-12b, WASP-18b, HAT-P-7b, OGLE-TR-56b, WASP-19b, CoRoT-1b, WASP-33b, HD149026b), suggesting that these atmospheres could exhibit different behaviors from the rest. 

Following this, \citet{Schwartz2015} calculate the dayside temperature of 50 planets with thermal eclipse measurements in at least two infrared wavelengths (>0.8~$\mu$m). They note an unexpectedly steep correlation, such that the hotter planets had temperatures even hotter than irradiation temperature predictions. This supports the previous claim by \citet{Cowan2011b} that the hottest planets have lower Bond albedo and/or less efficient heat transport. 
\citet{Schwartz2017} incorporate phase offsets into their energy budget calculations of six planets, which pushes the results toward lower Bond albedos and slightly higher heat transport than before. \citet{Keating2019} and \citet{Beatty2019} estimate the nightside temperature of several hot Jupiters using Spitzer phase curves and find that despite the different levels of irradiation, they all demonstrate similar nightside temperatures. This suggests that they might all have some chemically similar high optically thick cloud layer that is emitting at the nightside temperature.

Additionally, \citet{Garhart2020} perform  uniform analyses of 36 planets with Spitzer/IRAC secondary eclipses at 3.6 and 4.5~$\mu$m. They find an increasing trend in the brightness temperature ratio with equilibrium temperature. They find that this trend is present throughout the entire temperature range continuously between the coolest and the hottest planets (800K to 2500K). 

Our study builds on the previous works by expanding to 78 planets, with almost double the number of ultra-hot Jupiters, and by employing a careful treatment of the stellar flux. We use the two warm Spitzer/IRAC bandpasses (3.6~$\mu$m and 4.5~$\mu$m) \citep{Fazio2004, Werner2004} to study the near infrared trends in hot Jupiter emission. At these wavelengths, based on equilibrium chemistry, we expect to see spectral signatures of methane (CH4) (in the cooler planets) and carbon monoxide (CO) (in the hotter planets).
More specifically, we focus on the deviation of these points from a blackbody, particularly on its effect when including the ultra-hot Jupiters. Furthermore, we now compare our results to a grid of forward models that encompass the processes relevant for the coolest to the ultra-hot planet atmospheres (molecular dissociation, \ce{H-} opacity, latent heat, and the formation of temperature inversions). In Section \ref{P2:sec:obs} we describe the Spitzer/IRAC observations and data collection. In Section \ref{P2:sec:data} we describe the data analysis and the various temperatures used. In Section \ref{P2:sec:results} we present the results of the survey, we make a comparison to blackbodies, and demonstrate a transition to the ultra-hot Jupiters. In Section \ref{P2:sec:disc} we interpret our results in terms of albedo, redistribution, and temperature inversions.


\section{Observations}
\label{P2:sec:obs}

Our comprehensive survey is composed of 78 planets with eclipse depths taken with the Spitzer/IRAC at 3.6~$\mu$m and 4.5~$\mu$m. 
The literature data for the planets in this survey were collected via exoplanets.org \citep{Han2014}, exoplanet.eu \citep{Schneider2011}, or directly from the studies. We analyzed two 4.5~$\mu$m eclipses of KELT-9b \citep{Mansfield2020} using our custom pipeline (Baxter et al. in prep.) implementing Pixel Level Decorrelation to correct systematics \citep{Deming2015} (Appendix \ref{P2:app:kelt9b}). The planets, eclipse depths, stellar parameters, references, and key results and uncertainties are displayed in Table \ref{P2:tab:longtab}. 
Our work relies on the calculation of the equilibrium temperature, and since this parameter is sensitive to the eccentricity of the planetary orbit, especially on short period exoplanets, we  opted to perform an eccentricity cut and only select planets with eccentricity less than 0.2. 


\section{Data analysis}
\label{P2:sec:data}

\subsection{Calculating the planetary brightness temperatures}
\label{P2:app:Tbcalc}

The secondary eclipse depth measures the ratio of the planetary flux ($F_p$) to the stellar flux ($F_s$) at a given spectral bandpass. The planets selected for our survey have eclipse depths ($F_p/F_s$) measured in the two Spitzer/IRAC bandpasses (3.6~$\mu$m and 4.5~$\mu$m) \citep{Werner2004}. We remove the contribution of the stellar flux from the eclipse depths and convert the planetary flux to flux density ($\si{erg.cm^{-2}.s^{-1}.\AA^{-1}}$), which we use to calculate the brightness temperature by inverting the Planck function for the planet

\begin{equation}
    T_b (\lambda) = \frac{hc}{k_b\lambda} \Big[ \ln \left( \frac{2hc^2\pi \delta_{tra} }{\lambda^5 \overline{F}_{s}(\lambda) \delta_{occ} }   \right) \Big]^{-1},
    \label{P2:eq:Tb}
\end{equation}where $\delta_{tra}$ is the published transit depth $(R_p/R_s)^2$, and $\delta_{occ}$ is the eclipse depth measured at the Spitzer wavelengths $(F_p/F_s)$, $\lambda$ is the wavelength of the observed eclipse depth, either 3.6~$\mu$m or 4.5~$\mu$m, and $\overline{F}_{s}(\lambda)$ is the flux density of the stellar model weighted by the Spitzer/IRAC spectral response at this wavelength. 

Since both the planetary and the stellar model need to be integrated over the Spitzer spectral response functions, the spectral response weighted brightness temperature needs to be calculated iteratively. We create a grid of brightness temperatures around an estimated value (obtained from solving equation \ref{P2:eq:Tb} directly) and convert this to a grid of eclipse depths by convolving both the planetary blackbody function and the stellar models with the spectral responses. Our adopted brightness temperature is thus the one that produces the eclipse depth which best matches the data (lowest $\chi^2$). For this minimization we chose grids encompassing 200K around the calculated brightness temperatures, with step sizes of 2K, which is much smaller than the typical uncertainty of 100K. We then confirmed that we had reached a minimum $\chi^2$ for each planet. 

The integration of the spectral response with the model flux densities is done using the following equation: 
\begin{equation}
    \overline{F}(\lambda) = \frac{\int_0^\infty  F(\lambda) \lambda R(\lambda) d\lambda}{\int_0^\infty  \lambda R(\lambda) d\lambda}
.\end{equation}Here $R(\lambda)$ is the spectral response function at either 3.6~$\mu$m or 4.5~$\mu$m [e-/photon] taken from \citet{Quijada2004} and $F(\lambda)$ is the flux density of the planet or the star. The output, $\overline{F}(\lambda)$, is the average flux density that would be observed with Spitzer/IRAC. 

We decided to estimate the uncertainties on the adopted brightness temperatures by taking the minimum and maximum eclipse depth (based on the 1$\sigma$ uncertainty presented in Table \ref{P2:tab:longtab}) and propagating it through Equation \ref{P2:eq:Tb} to calculate a minimum and maximum brightness temperature. The 1$\sigma$ uncertainty on the brightness temperature is then the mean of these two deviations from the best fit brightness temperature. Since the Rayleigh-Jeans limit is for long wavelengths and high temperatures, the Rayleigh-Jeans formula cannot be simply used to calculate the uncertainties as the temperatures of the planets are overestimated. However, the formula can be used to get an estimate of the uncertainties provided the temperatures used in the propagation are those calculated using the full Planck function. Our method  estimates uncertainties that are equivalent to those calculated using the differentiated and propagated Rayleigh-Jeans law formula. 


We test different stellar models to correct the stellar flux contribution when calculating $T_B$ (see Appendix \ref{P2:app:StellarModels}). We then compare these temperatures to theoretical predictions for the zero albedo full redistribution equilibrium temperature $T_{eq,\textit{0}}$, irradiation temperature $T_0$, and maximum dayside temperature $T_{max}$. Throughout this paper we fit all correlations with an orthogonal distance regression (ODR), and obtain uncertainties by bootstrapping (see Appendix \ref{P2:app:fitting}).

\subsection{Extracting the planetary flux deviation from a blackbody}

We define a new metric that allows us to self-consistently compare how similar these planets are to blackbodies. We do this by first calculating the brightness temperature at 3.6~$\mu$m then we propagate this as a blackbody to 4.5~$\mu$m and recalculate the expected eclipse depth at 4.5~$\mu$m. We measure the deviation between this value and the actual 4.5~$\mu$m eclipse depth (Observed - Calculated) and call it the deviation from the blackbody (devBB). A positive deviation means that the 4.5~$\mu$m eclipse depth is greater than expected. Uncertainties are fully propagated from the uncertainties on the eclipse depths at 3.6 and 4.5~$\mu$m. Results are displayed in Table \ref{P2:tab:longtab}. Since devBB is the difference of flux ratios it is unitless, but for convenience we express it as the difference in percentages. We also note that using the Rayleigh-Jeans law, we can demonstrate that the deviation from the blackbody is equivalent to the normalized difference in the brightness temperatures. However, it has the advantage that it is derived directly from an observable quantity, the planet-to-star flux ratio. 

\subsection{Irradiation, equilibrium, effective, and max dayside temperatures definitions}
\label{P2:app:Temps}

Following \citet{Hansen2008}, we define the irradiation temperature ($T_0$) as the temperature of the planetary atmosphere at the  substellar point caused by the irradiation received from the host star at the distance of the planetary orbit $T_0 = T_{eff} \sqrt{R_*/a}$, where $T_{eff}$ is the stellar effective temperature, $R_*$ is the stellar radius, and $a$ is the semi-major axis of the orbit (assuming a circular orbit). 
The equilibrium temperature is another theoretical calculation that takes into account the albedo of the planet and the amount of redistribution over the planet's surface. The equilibrium temperature for isotropic (full) redistribution of incoming irradiation is thus defined as $T_{eq} = T_{eff}(1-A_B)^{1/4} \sqrt{R_*/2a}$, where $A_B$ is the planetary Bond albedo. When we take the Bond albedo to be zero and assume full redistribution, the equilibrium temperature can be written in terms of the irradiation temperature: $T_{eq,\textit{0}} = (1/4)^{1/4} T_0$. Subsequently, we define the disk integrated apparent maximum dayside temperature \citep{Schwartz2017} as the equilibrium temperature where the incoming radiation is immediately re-radiated (i.e., no redistribution: $T_{eq,max} = (2/3)^{1/4} T_0$). We do not expect any planets to have temperatures hotter than this as we do not expect any heat from contraction since most of these stars have ages $\gtrapprox$ 1 Gyr. Furthermore, since temperatures add to the fourth power, even planets with a substantially high internal temperature \citep[e.g.,][]{Thorngren2019} would be within the noise for this study. The uncertainties on these temperatures are calculated through full propagation of uncertainties from the stellar and orbital parameters.

The final temperature used in our analysis is the planetary effective temperature used in Appendix \ref{P2:app:schwartzcomparison}. We calculate the average brightness temperature, which we take as the error weighted mean of the two brightness temperatures, such that $<T_B> = (T_{b_{3.6}}/ \sigma_{3.6}^2 +  T_{b_{4.5}}/\sigma_{4.5}^2) / 2 $, where $T_{b_{\lambda}}$ is the brightness temperature at wavelength $\lambda$ and $\sigma_{\lambda}$ is the corresponding error on this measurement. $<T_B>$ is algebraically the same as $T_{eff}$ defined in \citet{Schwartz2015} and \citet{Cowan2011}. 

\subsection{Grid of forward emission models to interpret observations}
\label{P2:sec:models}
We utilize a new grid of cloud-free self-consistent radiative-convective thermochemical-equilibrium grid models, ScCHIMERA, originally developed and validated against analytical solutions and previously published brown dwarf models in \citet{Piskorz2018} and subsequently applied to the UHJ datasets presented in \citet{Arcangeli2018, Mansfield2018} and \citet{Kreidberg2018}. These new models are a successor to the \citet{Fortney2008} models.

\begin{figure*}
    \centering
    \includegraphics[width=\linewidth]{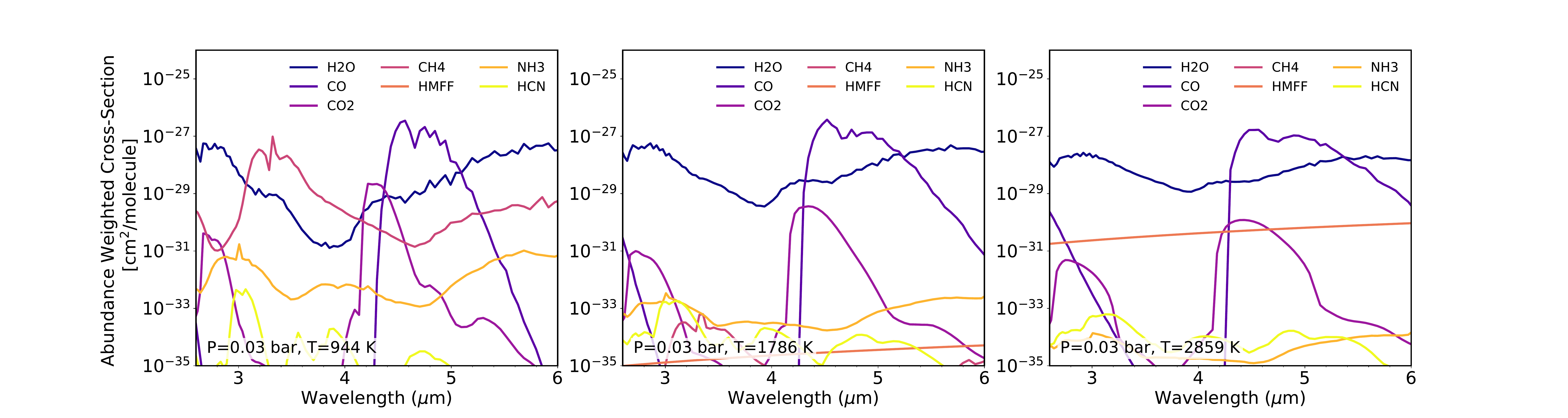}
    \caption{Abundance weighted cross sections for a selection of the emission model grid described in Section \ref{P2:sec:models}, \citep{Piskorz2018}. Each panel presents the abundance weighted cross sections for planets with equilibrium temperatures of 1000K, 1800K, and 3000K. Each TP profile is for a planet around a 5300K star with C/carbon to oxygen (O =) 0.54, [M/H] = 0.0, logg = 3.0; the full grid is shown in the first panel of Figure \ref{P2:fig:Linemodels}. The plotted abundances are taken from a pressure of 30mbar, resulting in probing temperatures of 944K, 1786K, and 2859K in each of the respective TP profiles.}
    \label{P2:fig:opacities}
\end{figure*}

Briefly, the model solves for the temperature profile through a vertical flux divergence minimizing via the Newton-Raphson iteration \citep{McKay1989} utilizing the two stream source function technique for the radiative fluxes \citep{Toon1989}. Mixing length theory is used to compute the convective fluxes as prescribed in \citet{Hubeny2017}.  Opacities (at R=100, 0.3 - 200~$\mu$m, where available) are treated within the correlated-K resort-rebin mixing framework \citep{Lacis1991, Amundsen2017} and include hot Jupiter-to-UHJ relevant atoms/molecules/ions: \ce{H2}-\ce{H2}/\ce{He} collision induced absorption, \ce{H2O}, \ce{CO}, \ce{CO2}, \ce{CH4}, \ce{NH3}, \ce{H2S}, \ce{HCN}, \ce{C2H2}, \ce{Na}, \ce{K}, \ce{TiO}, \ce{VO}, \ce{FeH}, \ce{H-} free-free/bound-free, \ce{PH3}, \ce{Fe}, \ce{Fe+}, \ce{Ca}, and \ce{Mg}, obtained from a variety of sources (ExoMol, \citet{Freedman2008, Freedman2014} and \citet{Kurucz1995}). Figure \ref{P2:fig:opacities} demonstrates a selection of the abundance weighted opacities extracted at the approximate pressure of the Spitzer contribution functions for three example planets (1000K, 1800K, and 3000K). Atom/Molecule/Ion abundances are computed using the Gibbs free energy minimization routine, NASA CEA2 \citep{Gordon1994}, given the specified scaling to the \citet{Lodders2009} elemental abundances. This approach also accounts for vertically varying abundances from thermal dissociation. The model assumes full redistribution at a given irradiation temperature, and an internal temperature of 150K (however, see \citet{Thorngren2019}). We utilize the PHOENIX \citep{Allard2011} models derived from the STScI pysynphot routine for the incident stellar flux (assuming a hemispheric mean incident flux--u=0.5).

\begin{figure*}
    \centering
    \includegraphics[width=\linewidth]{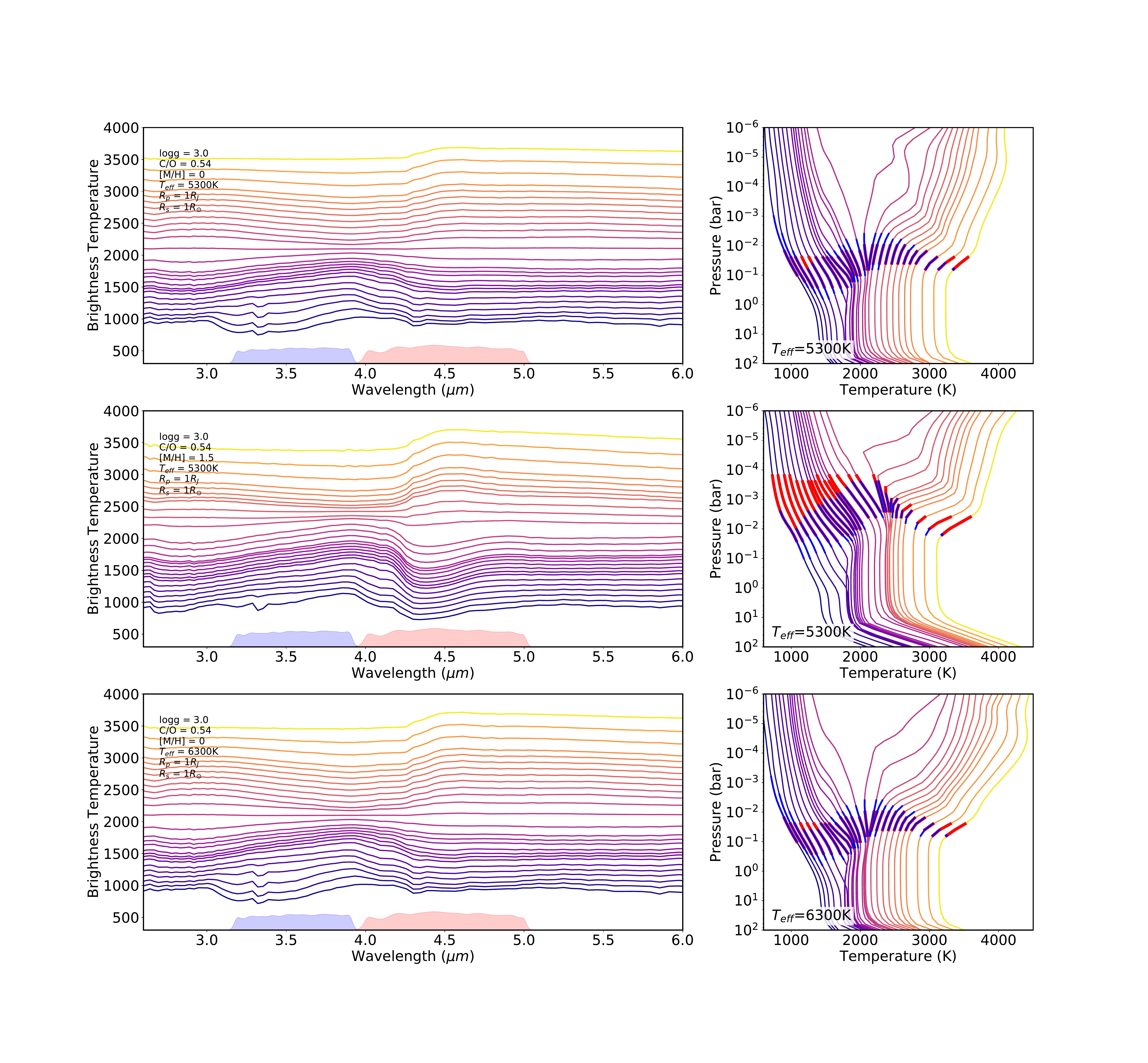}
    \caption{ScCHIMERA model emission spectra for hot Jupiters \citep{Piskorz2018} for a set of models of varying equilibrium temperature with log(g) = 3.0, C/O = 0.54, [M/H] = 0, $R_p$ = 1$R_J$, and $R_*$ = 1$R_\odot$. In each row we show the flux ratio (left) and temperature pressure profiles (right) for the 1D atmospheres of planets with colors indicating the increasing equilibrium temperatures ranging from 1000K to 3600K (in 100K increments). Top, middle, and bottom rows show the grid for planets around a 4300K, 5300K, and a 6300K star, respectively. Blue and red shaded regions in the left panel indicate the Spitzer/IRAC 3.6~$\mu$m and 4.5~$\mu$m bandpasses, respectively. Blue and red bold lines on the TP profiles correspond to the FWHM of the weighting functions for the 3.6~$\mu$m and 4.5~$\mu$m channels.}
    \label{P2:fig:Linemodels}
\end{figure*}

The model grid consists of 297 spectra and spans a range of carbon to oxygen ratios (C/O = 0.1, 0.54, 0.84), planetary surface gravities (log(g) = 2.5, 3.0, 3.5, 4.0 ), metallicity ([M/H] = -1, 0, 1, 1.5), stellar temperatures ($T_{eff}$ = 4300, 5300, 6300K), and planetary equilibrium temperatures ($T_{eq,\textit{0}}$ = 1000-3600K in steps of 100K).
Figure \ref{P2:fig:Linemodels} demonstrates a selection of the emission spectra at the Spitzer wavelengths, where $F_p/F_s$ is calculated using $R_s/R_p$ = 9.95. We show three tracks corresponding to the three different stellar temperatures. For 4300K and 6300K we fix [M/H] = 0, C/O = 0.54, and log(g) = 3, whereas for 5300K we show [M/H] = 1.5, C/O = 0.54, and log(g) = 3. The right panel contains the temperature pressure profile, which shows the atmosphere turning isothermal very briefly for planetary atmospheres with an equilibrium temperature of 1900K and the temperature inversion appearing for models with equilibrium temperature of 2200K. The left panel demonstrates the emission spectra: carbon monoxide can be seen clearly in emission for the hottest temperatures where the inversion exists.

\section{Results}
\label{P2:sec:results}

\subsection{Deviation between equilibrium and brightness temperatures}
\label{P2:sec:individualTb}

\begin{figure}
    \centering
    \includegraphics[trim={0cm 0cm 0cm 0cm},clip,width=\linewidth]{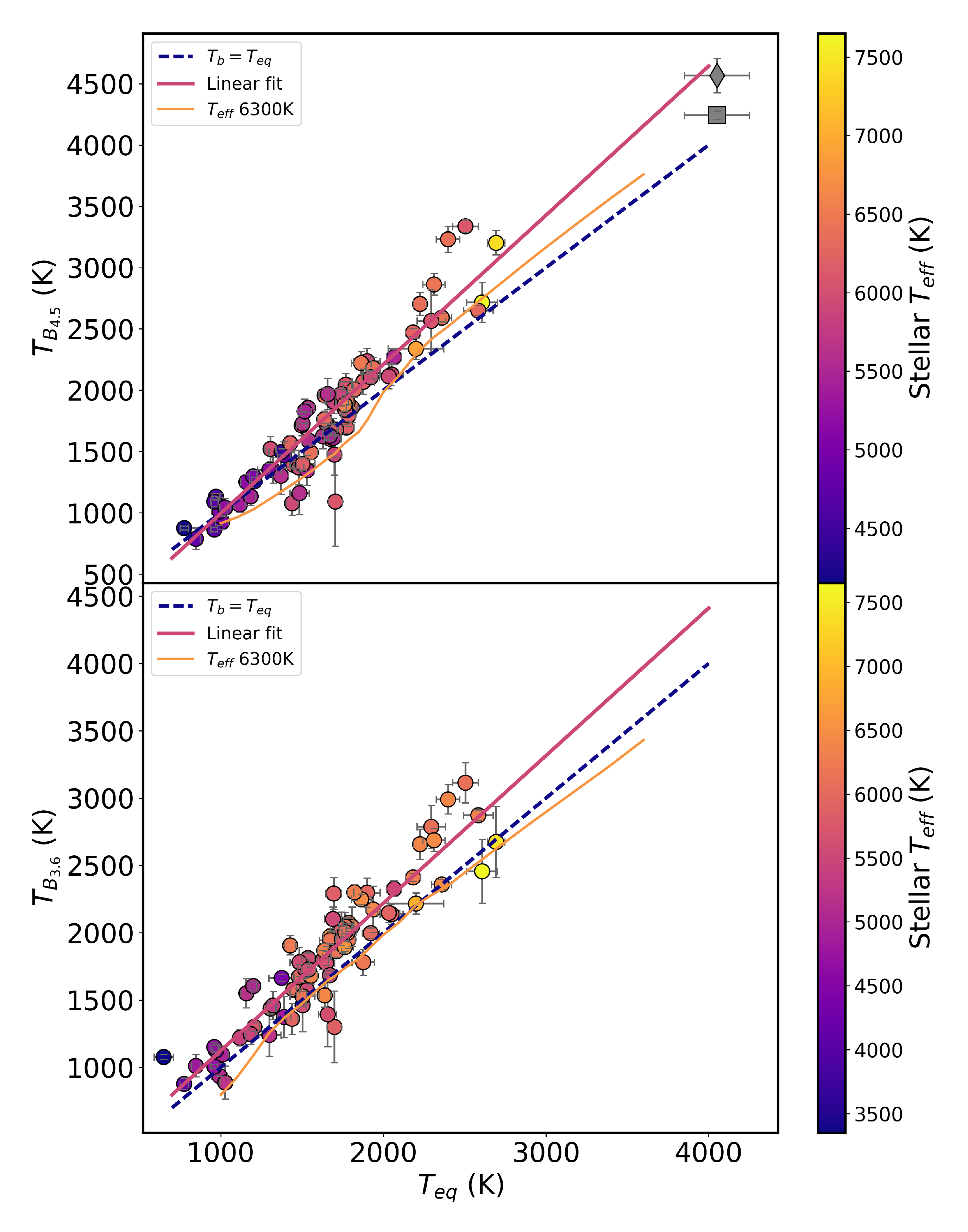}
    \caption{Brightness temperatures vs $T_{eq,\textit{0}}$ (full redistribution, 0 albedo) at 3.6~$\mu$m (bottom panel) and 4.5~$\mu$m (top panel). Magenta trend lines show a linear ODR fit to the data (gradient in the legends) and the blue dashed line shows the $T_B = T_{eq,\textit{0}}$ slope (gradient of 1). The gray points are the 4.5~$\mu$m brightness temperatures of KELT-9b: the square is our analysis presented in Appendix \ref{P2:app:kelt9b}, and the diamond is the analysis presented in \citet{Mansfield2020}. Forward ScCHIMERA models are displayed in orange for one stellar effective temperature of 6300K}.
    \label{P2:fig:Tbs}
\end{figure}

In Figure \ref{P2:fig:Tbs} we present the measured brightness temperatures plotted against $T_{eq,\textit{0}}$ for the two IRAC bandpasses. We fit linear functions using an orthogonal distance regression (ODR), see Appendix \ref{P2:app:fitting}). If the brightness temperature is the same as $T_{eq,\textit{0}}$ then the gradient of the slope will be unity. The measured gradients at 3.6~$\mu$m and 4.5~$\mu$m are $1.09\pm0.06$ and $1.19\pm0.06$, respectively. At 4.5~$\mu$m, this is a statistically significant (3.2$\sigma$) deviation from $T_{eq,\textit{0}}$. On the other hand, at 3.6~$\mu$m the brightness temperatures are consistent with the equilibrium temperature (1.5$\sigma$). Thus, the source of this deviation  exhibits a stronger effect at 4.5~$\mu$m compared to 3.6~$\mu$m. 
Furthermore, comparing to the grid of forward models demonstrates that the different stellar temperature model grids converge at lower temperatures and diverge at higher temperatures. We measure the residuals and standard deviations of the brightness temperatures to the best fit lines in three equally spaced temperature regimes (649K-1330K, 1330K-2012K, 2012K-2693K). At 4.5~$\mu$m the standard deviation of the residuals is 83K, 193K, and 258K, respectively,  for the low, medium, and high temperature bins. At 3.6~$\mu$m they are 157K, 187K, and 242K. The standard deviation of the residuals increases with increasing temperature, following the the trends predicted by the models with temperature inversions in Figure \ref{P2:fig:Tbs}.

Despite doing an eccentricity cut at an eccentricity of 0.2, there are still some planets with a nonzero eccentricity; these planets are typically outliers in Figure \ref{P2:fig:Tbs}. Eccentric orbits result in stellar insolation changing throughout the planets orbit, and thus it is expected that their measured brightness temperatures deviate from standard equilibrium temperature calculations (which assumes a circular orbit).

\subsection{Increasing trend in brightness temperature ratio versus equilibrium temperature}
\label{P2:sec:Tbratio}

\begin{figure}
    \centering
    \includegraphics[width=\linewidth]{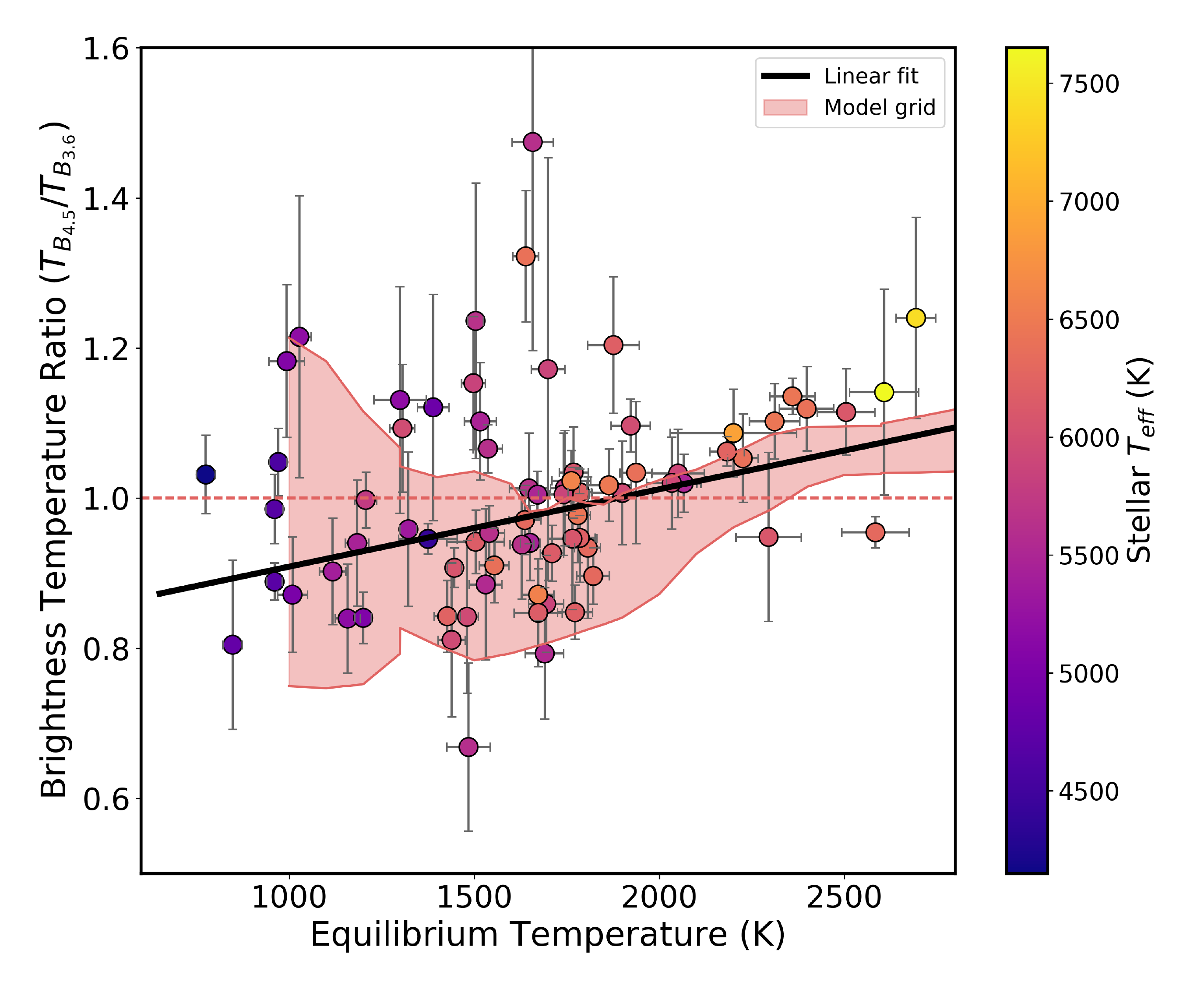}
    \caption{Brightness temperature ratio ($T_{B_{4.5}}/T_{B_{3.6}}$) vs the equilibrium temperature ($T_{eq,\textit{0}}$) of all of the available planets with secondary eclipses measured with Spitzer/IRAC. The blue line shows an ODR fit to the data with a slope significance of 3.1$\sigma$. Several functions were tested (see Section \ref{P2:sec:Tbratio}) and  the model with the lowest BIC is plotted as  a straight line. The orange shaded area shows the span of the ScCHIMERA model grid described in Section \ref{P2:sec:models}. The color scale is the effective temperature of the star.}
    \label{P2:fig:ratios}
\end{figure}

We demonstrate an increasing trend in the brightness temperature ratio with the $T_{eq,\textit{0}}$ (Figure \ref{P2:fig:ratios}). This is a manifestation of the 4.5~$\mu$m individual brightness temperatures deviating more from equilibrium than 3.6~$\mu$m as seen in Figure \ref{P2:fig:Tbs}. We fit the increasing trend, and find a slope of $95\pm31$~ppm per Kelvin for the PHOENIX models, which is significant to 3.1 $\sigma$. In addition to the linear fit we also test a bilinear model, but we find that the change in the BIC does not favor this scenario. Finally, we make a comparison with our grid of forward models and find that they are consistent with the data. 

\subsection{Increasing trend in planetary deviation from a blackbody}
\label{P2:sec:resultsdevBB}

\begin{figure}
    \centering
    \includegraphics[trim={0cm 0cm 0cm 0cm},clip,width=\linewidth]{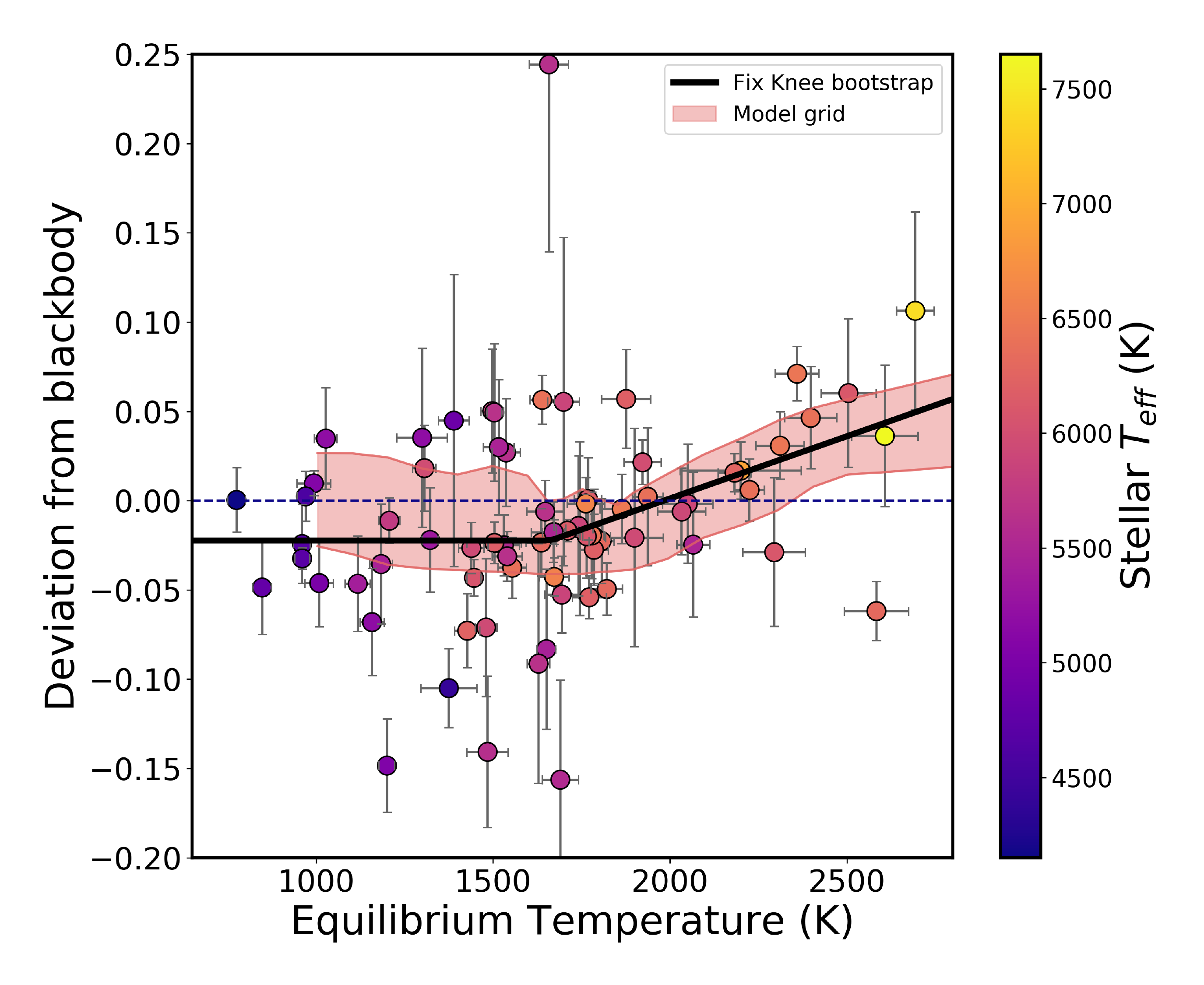}
    \caption{Deviation of the 4.5~$\mu$m eclipse depth from the 3.6~$\mu$m blackbody propagated to 4.5~$\mu$m vs equilibrium temperature (computed with zero albedo and full redistribution). Several functions were tested (see Section \ref{P2:sec:resultsdevBB}) and  the model with the lowest BIC  is plotted as a bilinear with a knee. The color bar presents the stellar effective temperature. The dashed horizontal line indicates a zero deviation, meaning that the eclipses are consistent with a blackbody. The orange shaded area represents the span of the fiducial forward ScCHIMERA models described in Section \ref{P2:sec:models}.} 
    \label{P2:fig:devBB}
\end{figure}

Assuming that the planetary flux is a blackbody set at 3.6~$\mu$m, we calculate the predicted eclipse depth at 4.5~$\mu$m and then calculate the deviation from the measured eclipse depth. Figure \ref{P2:fig:devBB} presents this deviation as a function of the equilibrium temperature. We fit three different trend lines to the data and compare their Bayesian information criteria (BIC). First, we fit a simple linear function (two free parameters), then we fit a bilinear model (four free parameters), and finally a bilinear model with the slope of the first line segment fixed to zero (three free parameters). The $\chi^2_{red}$ for the three models are 3.77, 3.51, and 3.50 and the BICs are 279, 261, and 259 for the straight line, bilinear, and fix bilinear, respectively. According to the $\Delta$BIC and $\chi^2_{red}$, the fixed bilinear model provides the best fit. This model captures a transition to the UHJs with an intercept of $1660\pm100$K. We also show that the grid of emission models are consistent with the data and predict these trends.


\section{Discussion}
\label{P2:sec:disc}

\subsection{Summary of our main results}

The 3.6~$\mu$m brightness temperatures are statistically consistent with $T_{eq,\textit{0}}$ (Figure \ref{P2:fig:Tbs}), but 4.5~$\mu$m shows a statistically significant increase in $T_B$ compared to $T_{eq,\textit{0}}$, which is seen as a continuum between the hot and ultra-hot planets. 
Additionally, the $T_{B_{4.5}}/T_{B_{3.6}}$ ratio also demonstrates a smooth continuous increase with $T_{eq,\textit{0}}$ (Figure \ref{P2:fig:ratios}). 
We note that, with our larger sample size, different uncertainty calculation, and different stellar model correction, we support the results of the linear fit of the $T_{B_{4.5}}/T_{B_{3.6}}$ ratio in \citet{Garhart2020} to better than $0.3\sigma$. 

However, in addition to the metrics in previous studies, our work includes the deviation from a blackbody which shows evidence of a transition between the hot and the ultra-hot Jupiters (Figure \ref{P2:fig:devBB}) that is not captured in the brightness temperature ratio (Figure \ref{P2:fig:ratios}). The deviation from a blackbody is proportional to the difference between the two brightness temperatures, whereas \citet{Garhart2020} present the ratio of the brightness temperatures. The ratio of two constantly increasing values is also a constant, but their difference is not. This subtle mathematical difference between the two metrics is the reason why a transition is not captured by the brightness temperature ratio.
A bilinear fit of the deviation from a blackbody is statistically favored, indicating that the  UHJs are driving the transition. This transition is also captured in our new grid of  1D self-consistent models (see Section \ref{P2:sec:modeldiscussion}). The 3.6 and 4.5~$\mu$m phase curve results of 12 hot Jupiters presented in \citet{Keating2019} tentatively support this transition in thermal structure. They visually demonstrate a difference in the temperature structures between the coolest and the hottest planets by plotting the difference in the two dayside brightness temperatures. 
We interpret below these trends and transitions in terms of temperature inversions and efficiency of redistribution. 

\subsection{Expected opacities at 3.6~$\mu$m and 4.5~$\mu$m}

The dominant absorbers at the wavelengths probed by Spitzer/IRAC are methane (3.6~$\mu$m), carbon monoxide (4.5~$\mu$m), and water (both wavelengths). \citet{Parmentier2018} and \citet{Lee2012} provide temperature--pressure profiles and the corresponding contribution functions for their analysis of emission spectra of WASP-121b (2400~K) and HD189733b (1200~K). Despite the different temperature regimes, the 4.5~$\mu$m contribution function probes lower pressures than 3.6~$\mu$m. This is driven by the bimodality at 4.5~$\mu$m caused by the \ce{H2O} deeper in the atmosphere ($\sim$30~mbar) and higher  \ce{CO}/\ce{CO2}  at lower pressures ($\sim$2-3~mbar), whereas 3.6~$\mu$m probes $\sim$40~mbar. 

The transition between the dominating carbon-bearing species in hot Jupiters is expected to occur at around 1000K \citep[e.g.,][]{Zahnle2014, Ebbing2016, Molaverdikhani2019}, with hotter atmospheres becoming dominated by CO. Consequently, any changes in the structure of the T-P profile would be seen at 4.5~$\mu$m due to the presence of CO \citep{Fortney2008, Parmentier2018, Arcangeli2018}. Specifically, a temperature inversion would result in CO in emission, increasing the 4.5~$\mu$m brightness temperature compared to 3.6~$\mu$m. As the planets approach the ultra-hot temperature regime, water and most other molecular species should begin to dissociate, except the CO. This will further increase the difference in the two pressures probed by Spitzer, making our observations even more sensitive to possible temperature inversions.

More generally, the peak of the Planck function corresponding to the thermal emission of the planet shifts at shorter wavelengths when the effective temperature of the planet dayside increases. Since the opacities generally increase with increasing wavelength, the difference between the opacities at the continuum and either of the Spitzer wavelengths then increases with increasing equilibrium temperature. The overall opacity of an atmosphere increases from $\sim1$~$\mu$m to 10~$\mu$m, mostly due to water. Therefore, any difference (positive or negative) between our measured $T_B$ and $T_{eq,\textit{0}}$ will be larger for hotter planets, as demonstrated in Figure \ref{P2:fig:Tbs}. However, the relative difference in the water opacity between the two Spitzer wavelengths is small enough that we do not expect the Planck function shift to be playing a role when comparing the brightness temperatures to each other (e.g., Figures \ref{P2:fig:ratios} and \ref{P2:fig:devBB}). Differences between the two Spitzer wavelengths are dominated by the CO opacity at 4.5~$\mu$m.

\subsection{Grid of forward models} 
\label{P2:sec:modeldiscussion}

In Figure \ref{P2:fig:Linemodels} we plot the range of 1D models from the emission model grid for three different stellar temperatures  (4300K top row, 5300K middle row, and 6300K bottom row). We can see that, for each model track, as the equilibrium temperature increases, the atmosphere switches from being non-inverted to being inverted. This causes the strong CO emission feature in the 4.5~$\mu$m bandpass to emerge. The hotter the equilibrium temperature of the planet, the stronger the temperature inversion, and the stronger the CO emission feature. We note that we also see the \ce{CH4} absorption feature appearing as a dip in the brightness temperatures at 3.6~$\mu$m for the coolest (non-inverted) models. The trend from hot to cold is from a weakening inversion until finally the TiO and VO condense out, with a very small isothermal transition region, as can be seen in the grid model T-P profiles displayed in Figure \ref{P2:fig:Linemodels}.

Additionally, Figure \ref{P2:fig:Linemodels} demonstrates that as the effective temperature of the star increases, the atmosphere of the planet with a given equilibrium temperature has a stronger inversion than a planet with the same temperature does around a cooler star. This is in part because at a given planetary temperature, the wavelength separation between the stellar spectrum and the planetary spectrum increases for hotter stars, which results in a higher effective visible-to-infrared Planck mean opacity. The atmosphere of the planet may respond differently to these fluxes, resulting in different temperature pressure profiles.

We compare the complete sample of eclipses to our grid of 1D emission models for the individual planetary brightness temperatures, for which a subset is plotted in  Figure \ref{P2:fig:Linemodels}. We highlight that since most of the hottest ($T_{eq,\textit{0}}>2000K$) planets in our sample have stellar temperatures $>5900K$ they should be modeled by the 6300K track. 
We plot modeled tracks corresponding to planets around a 6300K star on Figure \ref{P2:fig:Tbs}. We find that the temperatures we measured for our survey planets are higher than expected from the model tracks. We interpret this as being due to the model equilibrium temperature assuming full uniform redistribution, whereas these planets are likely tidally locked and thus will have hotter daysides. However, we do find that the models capture the stronger deviation between brightness and equilibrium temperatures at 4.5~$\mu$m compared to 3.6~$\mu$m for hotter planets.

We use the full grid of emission models (see Section~\ref{P2:sec:models}) for comparison with the deviation measured in channel~2 (4.5~$\mu$m) from the blackbody estimated from channel~1 (3.6~$\mu$m) and with the brightness temperature ratio (Figure \ref{P2:fig:devBB} and Figure \ref{P2:fig:ratios} respectively). First, we find that the model grid is consistent with both of the trends we measured from the data. The models show a clear transition at $\sim1700K$, which is consistent with the transition temperature we fit from the data in Figure \ref{P2:fig:devBB}. Second, the envelope of models do not show the same abrupt transition at $\sim1700K$ in the brightness temperature ratio (Figure \ref{P2:fig:ratios}) as they do in the deviation from the blackbody. Instead, they show a continuous increase with equilibrium temperature, with significant variations at the lower temperatures, which is in agreement with the data and the straight line we fit in Section \ref{P2:sec:Tbratio}.

We find that the spread in the models for both the deviation from the blackbody (Figure \ref{P2:fig:devBB}) and the brightness temperature ratio (Figure \ref{P2:fig:ratios}) is primarily caused by differences in metallicity and C/O ratio, with surface gravity and stellar temperature having little effect here. Thus, using the grids of different C/O ratios we are able to evaluate trends from the whole population. We find that we can rule out tracks with a high C/O ratio of 0.85 ($\Delta$BIC of $\sim$270), meaning that the population of hot Jupiters statistically favors low or solar C/O ratios (C/O$\leq0.54$). This means that high C/O planets are rare (C/O$\geq0.85$). 

\subsection{Interpretation of the transition from hot Jupiters to ultra-hot Jupiters}

\subsubsection{Assumptions on albedo, redistribution, clouds, and thermal structure}
\label{P2:sec:albedoredist}

We compute the equilibrium temperature ($T_{eq,\textit{0}}$) assuming full redistribution and null Bond albedo, see Section \ref{P2:app:Temps}. Changing these assumptions would have an effect on our results. A nonzero albedo would result in the predicted theoretical equilibrium temperature being lower than $T_{eq,\textit{0}}$, and relaxing the full redistribution assumption would increase the predicted equilibrium temperature toward $T_{eq,max}$ (no redistribution).
This likely explains the few cooler planets whose brightness temperatures are lower than the equilibrium temperature (Figure \ref{P2:fig:Tbs}). 

In Figures \ref{P2:fig:Tbs} and \ref{P2:fig:ratios}, we find a continuous increase in the brightness temperature with $T_{eq,\textit{0}}$, with the hottest planets being even hotter than the predicted equilibrium temperature, especially at 4.5~$\mu$m. Empirical estimates of the Bond albedo for hot Jupiters and ultra-hot Jupiters range from 0 to 0.3 \citep{Schwartz2015, Schwartz2017}. A nonzero albedo would statistically lower $T_{eq}$ below $T_{eq,\textit{0}}$, which in turn would strengthen the deviation seen. Furthermore, Figure \ref{P2:fig:Tbs} demonstrates that the increase in brightness temperatures with equilibrium temperature is also predicted by the models that assume zero albedo and full redistribution. Increasing the albedo in the models would also strengthen this deviation. We thus do not think our zero albedo assumption  changes these trends.

On the other hand, a lower redistribution efficiency for the hottest planets would increase their $T_{eq}$, resulting in hotter brightness temperatures. However, a compilation of Spitzer phase curves shows no trend with the difference of the phase curve offsets at the two Spitzer wavelengths \citep{Beatty2019, Zhang2018}. This provides no evidence for potential different redistribution in the two IRAC bandpasses, and we would thus expect the deviation to be equal at the two wavelengths; however,  this is not observed (Figure \ref{P2:fig:Tbs}). 
We hypothesize that a broader range of redistribution efficiencies for the hotter planets could explain the increasing scatter with increasing $T_{eq,\textit{0}}$ in Figure \ref{P2:fig:Tbs}. The degree to which hot Jupiters redistribute heat has been known to vary from planet to planet \citep{Showman2002, Cowan2007, Cowan2011b, Knutson2007, Showman2011}.

Figure \ref{P2:fig:devBB} shows a transition at $T_{eq}\simeq$1700K in the dayside emission of our sample of 78 hot Jupiters. We find a similar transition in the new model grid described in Section \ref{P2:sec:models}. Dayside clouds made of large particles could, in theory, equalize the brightness temperature at 3.6 and 4.5~$\mu$m; however, we do not think that a transition from cloudy to cloud-free  is a likely explanation for the trend seen in Figure \ref{P2:fig:devBB}. The main reason for this is that a large majority of hot Jupiters show very low geometric albedos in optical wavelengths, indicative of daysides that are not significantly dominated by cloud opacity. Second, if clouds made of small particles ($\sim1\mu$m) were dominating the opacity structure in the Spitzer bandpasses, they would be even more dominant in the Hubble Space Telescope wide field camera 3 (HST/WFC3) bandpass. However, the two emission spectra of hot Jupiters (not ultra-hot) taken with the HST/WFC3 and with a good enough signal to noise ratio (HD209458b and WASP-43b) show evidence of water absorption and not the blackbody emission  expected from a cloudy dayside~\citep{Line2016, Stevenson2014b}. Additionally, clouds composed of reflective species would create large shifts in the optical phase curves \citet{Shporer2015}. In searching for these large shifts in phase curves measured with the Kepler telescope, there is evidence that clouds could be present in only a tiny fraction of the dayside~\citep{Parmentier2016}. Based on this range of evidence, we consider it reasonable in this paper to model the daysides of planets with $T_{eq}$>1400K as being
cloud-free.

We conclude that the main cause of the increase in brightness temperature with equilibrium temperature and of the increasing deviation from a blackbody is indeed physical, and is not due to our assumptions of the albedo, redistribution, or cloud-free atmosphere when calculating the equilibrium temperature.

\subsubsection{Transition in thermal inversions}

The strength of the deviation from blackbody calculation (Figure \ref{P2:fig:devBB}) is that it is free of redistribution and albedo assumptions; it simply compares 4.5~$\mu$m to 3.6~$\mu$m. Theoretically, the positive deviation from blackbody could be emission by CO at 4.5~$\mu$m (inverted T-P profile) or absorption by methane at 3.6~$\mu$m (non-inverted T-P profile). However, given equilibrium chemistry, methane is very unlikely to be in the hottest atmospheres. Moreover, three of our hottest planets have already been shown to have temperature inversions: WASP-33b, WASP-121b, and WASP-18b \citep{Haynes2015, vonEssen2015, Evans2017, Arcangeli2018, Kreidberg2018}. Furthermore, the best fitting model is bilinear with an intercept of $1660\pm100K$, highlighting the statistical power of the UHJ deviation. Our grid of forward models also predict a curve that is similar to this bilinear fit, capturing the location of the intercept of the two lines at $\sim$1700K.

Interestingly, this corresponds to the condensation temperatures of TiO and VO, which could be the origins of thermal inversions \citep{Hubeny2003, Burrows2007, Fortney2008}. We thus interpret that this deviation represents the transition to a different physical realm in these atmospheres, for example as the temperature approaches that of the UHJs, atmospheres transition from  non-inverted to inverted. For the cooler hot Jupiters, temperature inversions are suggested to be caused by the absorption of optical incoming stellar irradiation by gas phase TiO and VO \citep{Hubeny2003, Fortney2008}. On the other hand, for UHJs, inversions can form through other absorbers such as Na/FeH/Fe/Fe+/Mg \citep[e.g][]{Lothringer2018, Pino2020} or from lack of cooling due to molecular dissociation \citep{Parmentier2018}. As molecular dissociation occurs, \ce{H-} becomes an important opacity source, leading to blackbody-like emission spectra, as seen in HST/WFC3 near 1.4~$\mu$m \citep[e.g.,][]{Arcangeli2018}. WASP-12b is the biggest outlier in Figure \ref{P2:fig:devBB} (it has the lowest deviation from a blackbody for planets with $T_{eq}$\,$>$\,2500), but this planet is thought to have potential mass loss, and so our considerations may not apply to it directly \citep{Cowan2012, Bell2019}. 

In Figure \ref{P2:fig:Tbs} we observe a stronger deviation from equilibrium temperature at 4.5~$\mu$m compared to 3.6~$\mu$m. We interpret that at 4.5~$\mu$m we see CO in emission, whereas at 3.6~$\mu$m there is a weaker emission feature from non-dissociated \ce{H2O} originating deeper and cooler in the atmosphere. This is also captured by the grid of models, especially for the hot stars which represent the majority of the deviating planets. However, in Figure \ref{P2:fig:Tbs} there is a systematic discrepancy between the models and the data which is not not captured in Figure \ref{P2:fig:devBB} (i.e., our fitted lines lie lower than the models predict). Our interpretation for this discrepancy and for the intrinsic scatter of the brightness temperatures is that it is either due to the difference in strength of the inversions or that the models are not capturing all of
the underlying physics. For example, these models do not account for  atmospheric drag \citep[e.g.,][]{Arcangeli2019} or assume that stratospheres are cloud-free. Moreover, whatever the effect is, it does not appear to correlate uniquely with stellar insolation since planets with similar equilibrium temperatures can exhibit different strengths of deviation.

KELT-9b is the hottest known transiting exoplanet and is thus a great probe of the extreme scenarios that we have already discussed above. In Appendix \ref{P2:app:kelt9b} we measure the 4.5~$\mu$m eclipse depth of KELT-9b from an observation centered around eclipse and lasting three times the eclipse duration. We compare this with the results of \citet{Mansfield2020} who use the full phase curve observation. Our brightness temperature is 4.6$\sigma$ lower than the value calculated in \citet{Mansfield2020}, which is likely due to the underestimation of the eclipse depth in our modeling since we approximate the concave down phase variation with a linear function; this discrepancy has been studied before \citep[e.g.,][]{Bell2019}. Nevertheless, we plot both $T_B$ on Figure \ref{P2:fig:Tbs} and find that both follow the trend of increasing $T_B$ with $T_{eq}$. In particular, the brightness temperatures calculated from the phase curve in \citet{Mansfield2020} agree with our fitted trend line to <1$\sigma$. However, both brightness temperature calculations are cooler than an extrapolation of the model grid might suggest.  We hypothesize that this is due to possible partial CO dissociation, given the ultra-hot equilibrium temperature of KELT-9b, resulting in lower CO emission in the 4.5~$\mu$m dayside observation \citep[e.g.,][]{Kitzmann2018, Lothringer2018}. A dedicated modeling analysis would be necessary to confirm this hypothesis, which is beyond the scope of this work.

In summary, our work demonstrates that a transition exists in the infrared emission spectra between hot Jupiters and ultra-hot Jupiters and that this is likely due to a change between non-inverted and inverted temperature-pressure profiles as the stellar irradiation increases on these planets.

\section{Clues from HST/WFC3}
\label{P2:app:HST}

\begin{figure*}
    \centering
    \includegraphics[trim={3.5cm 1cm 0cm 1cm},clip,width=\textwidth]{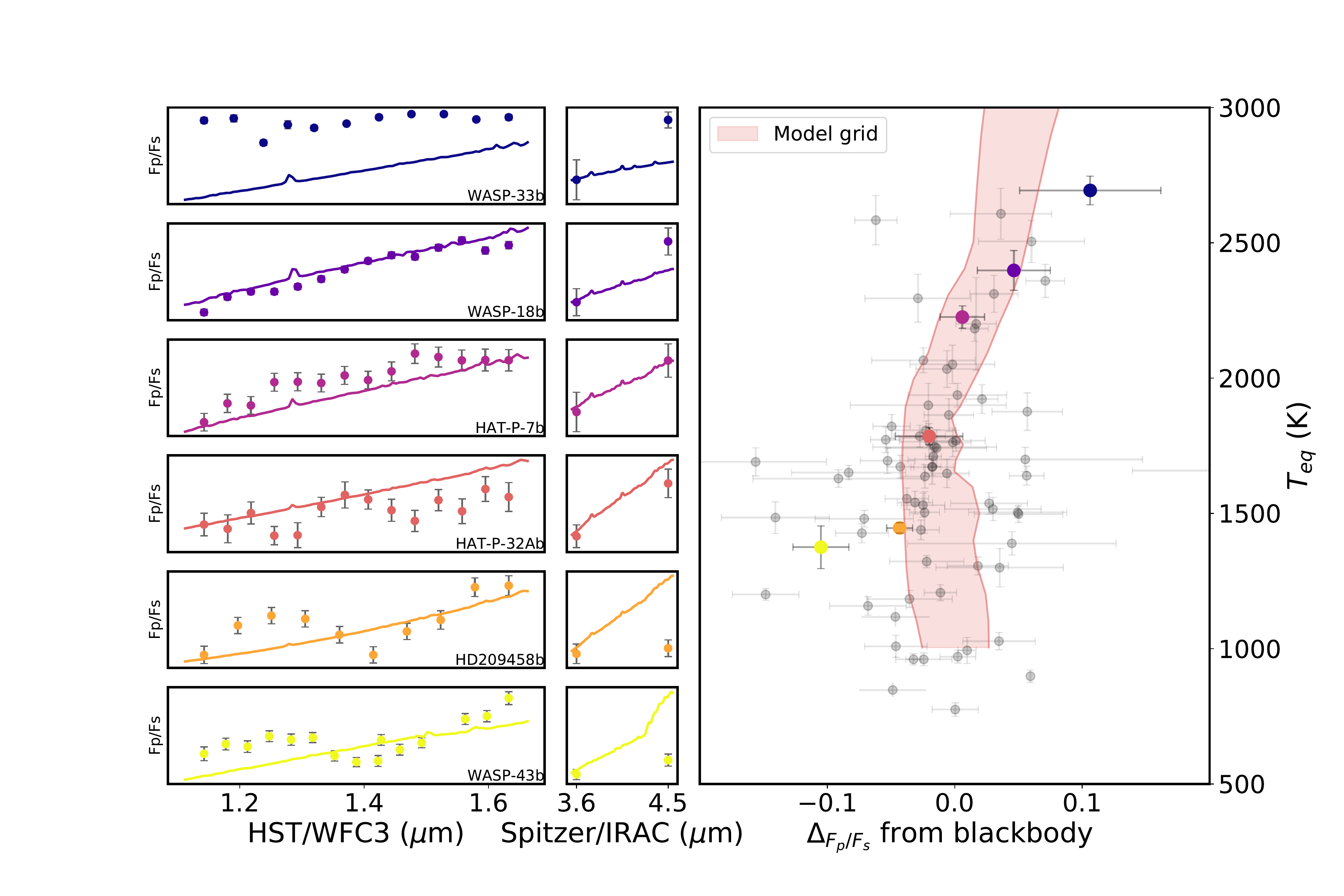}
    \caption{Right  panel: Deviation from the blackbody in the Spitzer bandpass against equilibrium temperature. Planets here demonstrate the continuous transition between the hot and the ultra-hot planets. Several planet with available HST spectra are highlighted and their spectra are plotted in the left   (HST/WFC3) and middle (Spitzer) panels. These planets are color-coded by increasing temperature. For simplicity and clarity, we show only six of the HST spectra as examples. The models shown in the left and middle panels are the blackbody at $T_{b_{3.6}}$ and PHOENIX model ratio emission spectra. The model overplotted on the rightmost panel is the emission model grid described in Section \ref{P2:sec:models}.}
    \label{P2:fig:devBBHST}
\end{figure*}

\begin{figure}
    \centering
    \includegraphics[trim={0.5cm 0.5cm 0.5cm 0.5cm},clip,width=\linewidth]{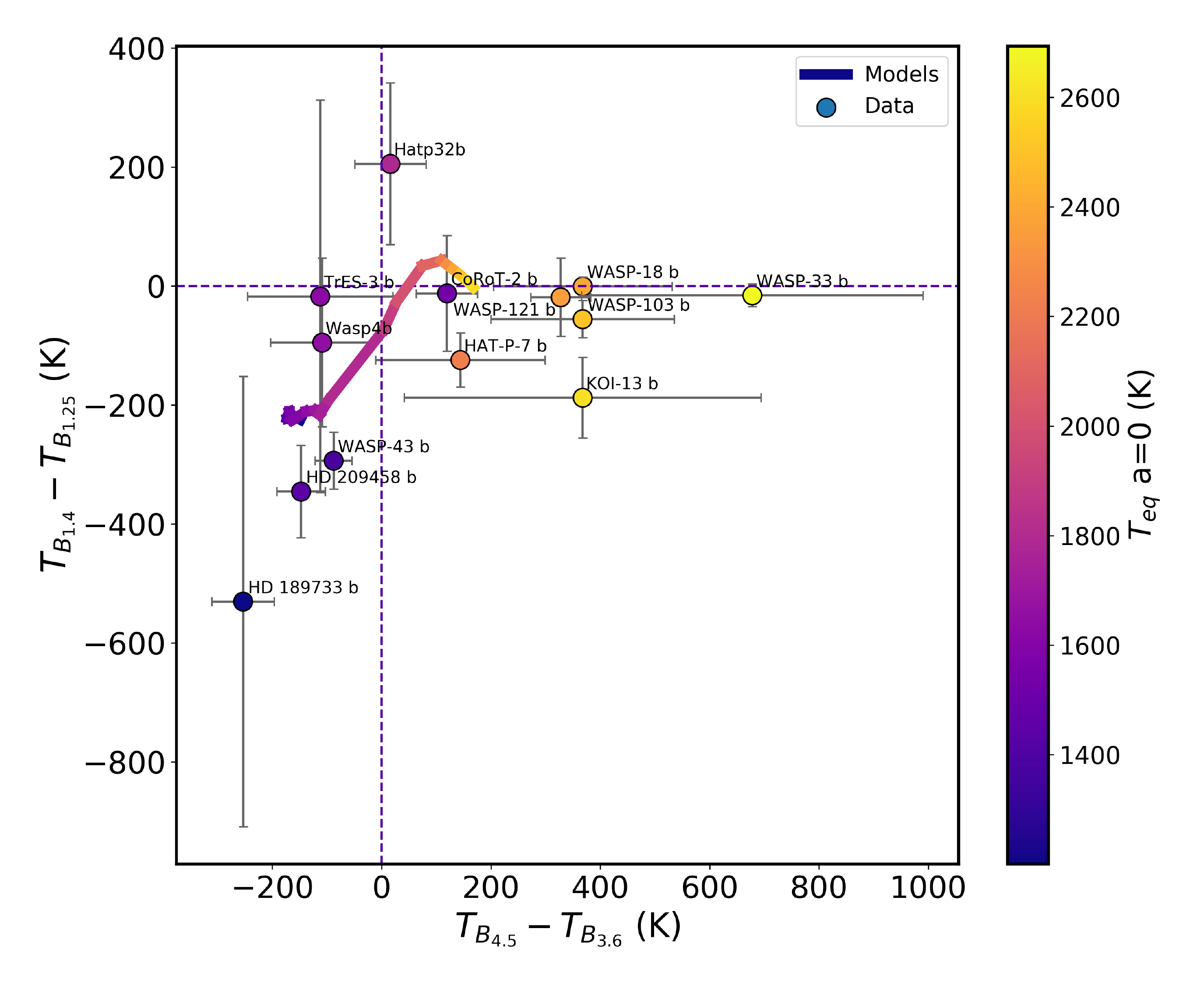}
    \caption{Color-color plot of the planets with available HST spectra. We calculate the color as the difference between the brightness temperatures in and out of the water feature in HST and as the difference between the brightness temperatures in the Spitzer bandpass.}
    \label{P2:fig:TbHST}
\end{figure}

Our knowledge of the physics occurring at the IRAC bandpass is deeply influenced by our knowledge of the spectrally resolved HST/WFC3 bandpass. We combine our Spitzer survey with available HST/WFC3 data from the literature and discuss the deviation from the blackbody in the context of the water feature at 1.4~$\mu$m in the HST/WFC3 spectral band. Figure \ref{P2:fig:devBBHST} shows the deviation from the blackbody calculated at the Spitzer wavelengths (right panel). This is combined with the individual Spitzer emission photometry and the HST emission spectra for a subsample of the available planets (middle and left panels respectively).

We see a continuum between the coolest and the hottest hot Jupiters. The hottest planets in our sample (WASP-33b and WASP-18b) have blackbody-like  spectra in HST/WFC3 caused by the \ce{H-} opacity \citep{Arcangeli2018}, and they show signs of a temperature inversion in Spitzer. HAT-P-32Ab is centered in the middle of the deviation from a blackbody plot, exhibiting no absorption or emission of CO, and shows a similar (albeit  noisier) blackbody emission spectrum with WFC3 \citep{Nikolov2018}. Finally, as we approach the coolest planets in HST (HD209458b and WASP-43b $\sim1500K$), we see the water feature appearing strongly in absorption. For these planets we see a negative deviation from the blackbody in Spitzer. We interpret this negative deviation as CO in absorption at 4.5~$\mu$m since at these cooler temperatures we expect to have non-inverted TP profiles. We highlight that our grid of models predicts these observations.

Building on the color-magnitude work of \citet{Triaud2014b} we also create a color-color diagram, where we use the difference between two brightness temperatures. Figure \ref{P2:fig:TbHST} shows the Spitzer color plotted against the HST color. The HST/WFC3 color is designed to capture inside and outside the water feature at 1.25~$\mu$m and 1.4~$\mu$m. We also show horizontal and vertical dashed lines representing blackbodies for each regime as well as the fiducial model track from our forward model grid ($T_{eff}$ = 5300K, C/O=0.54, [M/H] = 0, log(g)=3.0). Following the increasing temperature of the model track demonstrates the manifestation of the changing TP profiles (seen in Figure \ref{P2:fig:Linemodels}). The Spitzer color (horizontal axis), becomes larger as the models switch from exhibiting CO in absorption to CO in emission, whereas the HST color is slightly more complicated (vertical axis). First, there is a group of models around -200K (both axes) with negative colors, capturing the strong water absorption feature. This is followed by an increase toward a blackbody as the strength of the water feature decreases (-120K to +50K on the X-axis) and the atmospheres begin to transition toward thermally inverted with a slightly positive HST color, up to $\sim50$K. Finally, beyond a mid-IR color of +100K, the model HST colors become consistent with blackbodies again as the water feature disappears as the \ce{H-} opacity takes over.

In the available data, we note a clear gap in measured planetary temperatures (between HAT-P-7b (2225K) and HAT-P-32Ab (1785K)) where we expect to be probing the transition, which allows us to split the data into two families. The hotter sample planets ($>1785K$) have an average Spitzer color of 350K and exhibit less variance in the HST color, which captures the CO in emission at 4.5~$\mu$m and of their blackbody-like features in HST. Instead, the cooler sample planets ($\leq1785K$) have an average Spitzer color of $\sim$-80K, indicative of CO in absorption at 4.5~$\mu$m. Furthermore, the cooler sample follows the increasing model track as the strength of the water feature becomes less strong. Thus, our data   largely follow  the trends predicted by the models in both HST and Spitzer wavelengths, and we find that the published sample of HST data supports our claim of a continuum to the ultra-hot Jupiters. An analysis of an expanded dataset including new HST/WFC3 emission spectra for transiting giant planets will be presented in a forthcoming paper (Mansfield et al., in prep.).


\section{Conclusions}

We present our analysis of a literature survey of 78 hot Jupiters with secondary eclipses observed with Spitzer at 3.6~$\mu$m and 4.5~$\mu$m. Our survey spans equilibrium temperatures (zero albedo and full redistribution) between 800K and 2700K. We tested different stellar models (blackbody, ATLAS, PHOENIX) in order to correct the stellar flux from the secondary eclipse depths, and found that improper treatment of the star could bias results, particularly for planets around hotter stars. We then calculated the brightness temperatures at the two Spitzer wavelengths by using PHOENIX models to correct the stellar flux, by inverting the Planck function and integrating over the Spitzer spectral responses.

We find that the brightness temperatures at 4.5~$\mu$m are increasingly hotter than equilibrium temperature predictions for the hotter planets, which we interpret as a result of seeing CO in emission at 4.5~$\mu$m due to the temperature inversions in combination with the Planck function shift. The Planck function of a planetary atmosphere shifts to shorter wavelengths for higher temperatures, increasing the difference between the pressures probed by the equilibrium temperature and the pressures probed by Spitzer, and thus the magnitude of the difference between the brightness temperature and equilibrium temperature will be larger for hotter planets. However, we note that any differences between 3.6~$\mu$m and 4.5~$\mu$m are going to be dominated by the strong CO band at 4.5~$\mu$m.

We confirm a previous finding that the $T_{B_{4.5}}/T_{B_{3.6}}$ ratio exhibits a smooth continuum increasing with $T_{eq,\textit{0}}$.
However, we also measure, for the first time, the deviation of the data from the blackbody, which we defined as the difference between the observed 4.5~$\mu$m eclipse depth and the eclipse depth expected at 4.5~$\mu$m based on the brightness temperature measured at 3.6~$\mu$m. We find a transition at an equilibrium temperature of $1660\pm100$K in the deviation of the data from a blackbody. 

We compare our result to a new grid of 1D self-consistent models (ScCHIMERA) which contain the appropriate physics for temperature inversion formation. We find that the model grid is consistent with both of the trends we measured from the data; in particular, we find an excellent agreement between our measured transition and what is expected from the models. We suggest that this transition is capturing a change in the temperature pressure profile of these atmospheres, from non-inverted to inverted atmospheres as the stellar irradiation increases on these planets.

We find that the spread in the models for   the deviation from the blackbody and for the brightness temperature ratio is primarily caused by differences in metallicity and C/O ratio, with surface gravity and stellar temperature having little effect here. We rule out tracks with a high C/O ratio (0.85), meaning that the population of hot Jupiters statistically favors low or solar C/O ratios (C/O$\leq0.54$), and that high C/O planets are rare (C/O$\geq0.85$).



\begin{acknowledgements}

We thank the referee Nick Cowan, for his thorough work during the reviewing process. This work is based on observations made with the Spitzer Space Telescope, which is operated by the Jet Propulsion Laboratory, California Institute of Technology under a contract with NASA. Support for this work was provided by NASA through an award issued by JPL/Caltech. J.M.D acknowledges support from NASA grant NNX16AC64G, the Amsterdam Academic Alliance (AAA) Program, and the European Research Council (ERC) European Union’s Horizon 2020 research and innovation programme (grant agreement no. 679633; Exo-Atmos). 

\end{acknowledgements}


\bibliographystyle{aa}
\bibliography{bib2_out}

\begin{appendix}

\section{Details of the data analysis}

\subsection{Fitting correlations with x and y errors}
\label{P2:app:fitting}

Fitting of linear functions is often done using an ordinary least squares (OLS) or Markov chain Monte Carlo (MCMC) method, both of which assume Gaussian errors. However, our data has errors on both the abscissa and the ordinate, meaning a simple OLS cannot be performed \citep{Hogg2010}. We opted for the \textit{scipy.odr} package, translated from the FORTRAN-77 ODRPACK by \citet{Boggs1989}. ODRPACK is a weighted orthogonal distance regression function which takes into account errors on x and on y by minimizing the weighted orthogonal distance between the observations and the model. However, as pointed out in \citet{Beatty2019}, ODRPACK uses relative errors between the data points, meaning that the resulting covariance matrix remains the same even when you multiply all of the individual errors by some factor. This has the potential for producing incorrect uncertainties on the parameters. \citet{Beatty2019}  use another package, bivariate correlated errors and intrinsic scatter (BCES) \citep{Akritas1996, Nemmen2012}. However, this package only fits a linear model, and so is not suitable for our cases. 

Furthermore, these regression methods rely on the assumption that the model perfectly captures the data and that the data are drawn from a purely Gaussian distribution \citep[e.g.,][]{Galton1886, Zhang2004}. In our case, we know that both of these assumptions are not true, and that estimating errors from the covariance matrix could result in underestimated uncertainties. We thus decide to sample the parameter space using bootstrapping. Bootstrapping estimates posterior distributions by repeatedly resampling with replacement and refitting the function \citep{Efron1993}. We use ODR to fit the function, accounting for errors on x and y, and then we bootstrap to obtain parameter distributions. Our parameter estimates are then quoted as the 16th, 50th, and 84th percentiles on the marginalized parameter distributions.  

When measuring the slope of brightness temperature ratio against equilibrium temperature, we find a slope of $95\pm31$~ppm, which has a significance of 3.1$\sigma$, see Section \ref{P2:sec:Tbratio}. This is consistent but slightly less significant than 4$\sigma$ the result presented in \citet{Garhart2020} ($98\pm26$), despite our larger sample size. We test our method with their sample and still cannot reproduce their accuracy. We thus expect that the difference is simply due to the fitting and sampling methods chosen. \citet{Garhart2020} use a Gibbs MCMC sampler assuming Gaussian errors based on methods described in \citet{Kelly2007}, whereas our bootstrap method does not assume that the errors are Gaussian, and thus end up with broader posterior distributions for our parameters.

\section{Importance of using stellar models}
\label{P2:app:StellarModels}

The calculation of the brightness temperature requires an assumption of the stellar model in order to disentangle the planetary flux from the measured planet-to-star flux ratio ($F_p/F_s$). The simplest assumption is to model the star as a blackbody using the Planck function; however, it is also possible to use a grid of synthetic stellar models. For the first time, to our knowledge, we use our survey to test three different types of models for the star: blackbodies, ATLAS models \citep{Kurucz1979}, and PHOENIX models \citep{Allard1995, Husser2013}. 

For ATLAS models we use the ATLAS9 version of the code \citep{Castelli2003}. This assumes steady-state plane-parallel layers in local thermodynamic equilibrium (LTE), and opacities that are treated by averaging the contribution of different molecular and/or atomic species resulting in a line blanketing effect. Conversely, the PHOENIX models assume spherical geometry and direct opacity sampling of molecular and atomic species. They are also computed under the LTE assumption; however, non-local thermodynamic equilibrium (NLTE) effects are included for the spectral line profiles of selected important species (Li I, Na I, K I, Ca I, Ca II).

\subsection{Effect of different stellar models on measured temperature}

\begin{table}[]
    \caption{We measure the gradient of temperature vs irradiation temperatures for three different temperatures: the individual brightness temperatures ($T_{b_{3.6}}$ and $T_{b_{4.5}}$) and the dayside effective temperature ($T_{day}$), calculated as a weighted mean. Each set of temperatures is calculated using three different stellar models: blackbodies, ATLAS models, and PHOENIX models. Figure \ref{P2:fig:Tbs} displays the individual brightness temperatures for PHOENIX models and Figure \ref{P2:fig:Td} displays the effective temperatures.}
    \centering
    \begin{tabular}{l l l l }
    \hline\hline
          &   $T_{eff}$     & $T_{b_{3.6}}$ & $T_{b_{4.5}}$ \\
    \hline
    BB    &   0.85$\pm$0.03 & 0.81$\pm$0.05 & 0.92$\pm$0.05 \\
    ATLAS &   0.80$\pm$0.03 & 0.79$\pm$0.04 & 0.84$\pm$0.04\\
    PHOENIX & 0.80$\pm$0.03 & 0.76$\pm$0.05 & 0.84$\pm$0.05 \\
    \hline
    \end{tabular}

    \label{P2:tab:slopes}
\end{table}

Comparing the gradients (presented in Table \ref{P2:tab:slopes}) of each set of brightness and effective temperatures quantifies the difference between the stellar models. For the effective temperature we can see that the ATLAS and PHOENIX models are consistent with each other at better than $1\sigma$ level; however, blackbodies are systematically $\sim2\sigma$ above the stellar models. This larger gradient was also seen in \citet{Schwartz2015} with their sample, where they measured a value of 0.87(5) for the effective temperature. If any of these temperature sets were to be representative of the equilibrium temperature then the expected gradient would be 0.71 ($T_{eq} = (1/4)^{1/4}T_0 = 0.71 T_0$). The \citet{Schwartz2015} of 0.87(5) is statistically significantly steeper than 0.71, which they interpreted as hotter planets having a low Bond albedo and/or less efficient heat transport in their atmospheres. However, the gradient displayed in Table \ref{P2:tab:slopes} shows that this could also be an effect of the use of blackbodies to correct the stellar flux, and thus blackbodies cannot be excluded as the cause of their deviation. 

A similar result is seen in the individual brightness temperatures, whereby ATLAS and PHOENIX models are consistent with each other. Thus, for statistical studies of the planets with a wide range of temperatures, using blackbodies for the star can be misleading. We therefore decided to use stellar models to correct the stellar flux from eclipse measurements of our sample of hot and ultra-hot Juptiers. PHOENIX models have some advantages over other stellar models; they are computed at a higher resolution, span a larger range of temperatures, and contain direct opacity sampling. Additionally, PHOENIX models also account for some NLTE effects, which has been shown to be important for ultra-hot planets \citep{Lothringer2019}. Thus, we decided to use PHOENIX models instead of ATLAS for the remainder of the analysis. 

We also found that integrating over the spectral response increases the measured flux compared to taking the exact flux density at the average wavelength of the Spitzer band pass, as is done in \citet{Schwartz2015}. This is due to the nonlinear slope of both the stellar and planetary models over the bandpass, see Figure \ref{P2:app:StellarModels}. We calculate that ignoring this effect could lead to  planetary brightness temperatures overestimated by as much as 115K at 3.6~$\mu$m. This effect is more prominent where the slope of the spectra are steeper (e.g., at 3.6~$\mu$m compared to 4.5~$\mu$m or for hotter planets and stars). On average, for the whole sample, without integration we calculate overestimation of 32K at 3.6~$\mu$m and 14K at 4.5~$\mu$m. Thus, if not accounted for, this could enhance or diminish any statistical differences seen when comparing 3.6~$\mu$m and 4.5~$\mu$m. Additionally,  we find the hottest planets in our survey around the hotter stars. Thus, when looking for trends throughout a wide range of temperatures, it is imperative to carefully correct for the stellar flux  to ensure that what we are seeing is a result of the planetary atmosphere.

We find the trend in the brightness temperature ratio with equilibrium temperature to be $99\pm37$~ppm with blackbodies $97\pm35$~ppm with ATLAS models and $95\pm31$~ppm with PHOENIX models (Section \ref{P2:sec:Tbratio}. The maximum difference between the significance of these slopes for all of three stellar models is negligible, which suggests that any importance of stellar models  vanishes somewhat when looking at the ratios. Nevertheless, we keep the results from the PHOENIX models (Section \ref{P2:sec:Tbratio}).

\section{Comparing effective temperatures with \citet{Schwartz2015}} 
\label{P2:app:schwartzcomparison}

\citet{Schwartz2015} measure the slope of the effective dayside temperature against the irradiation temperature. They note that their slope $0.87\pm0.05$ is significantly steeper than equilibrium temperature predictions (0.71), and that this increasing deviation could  lower redistribution efficiencies in the hottest planets. We recreate their results with our expanded survey. We follow their method for calculating the effective temperature, which is the weighted mean of the brightness temperatures, and thus we call it $<T_B> = (T_{b_{3.6}}/ \sigma_{3.6}^2 +  T_{b_{4.5}}/\sigma_{4.5}^2) / 2 $. We then fit the resulting trends with an ODR (see Section \ref{P2:app:fitting}. However, first we test their method of brightness temperature calculation of inverting the Planck function and using a blackbody for the star. Then we test our method using a stellar model and fully integrating the Planck function. 

\begin{figure}
    \centering
    \includegraphics[trim={0.5cm 0cm 0.5cm 0cm},clip,width=\linewidth]{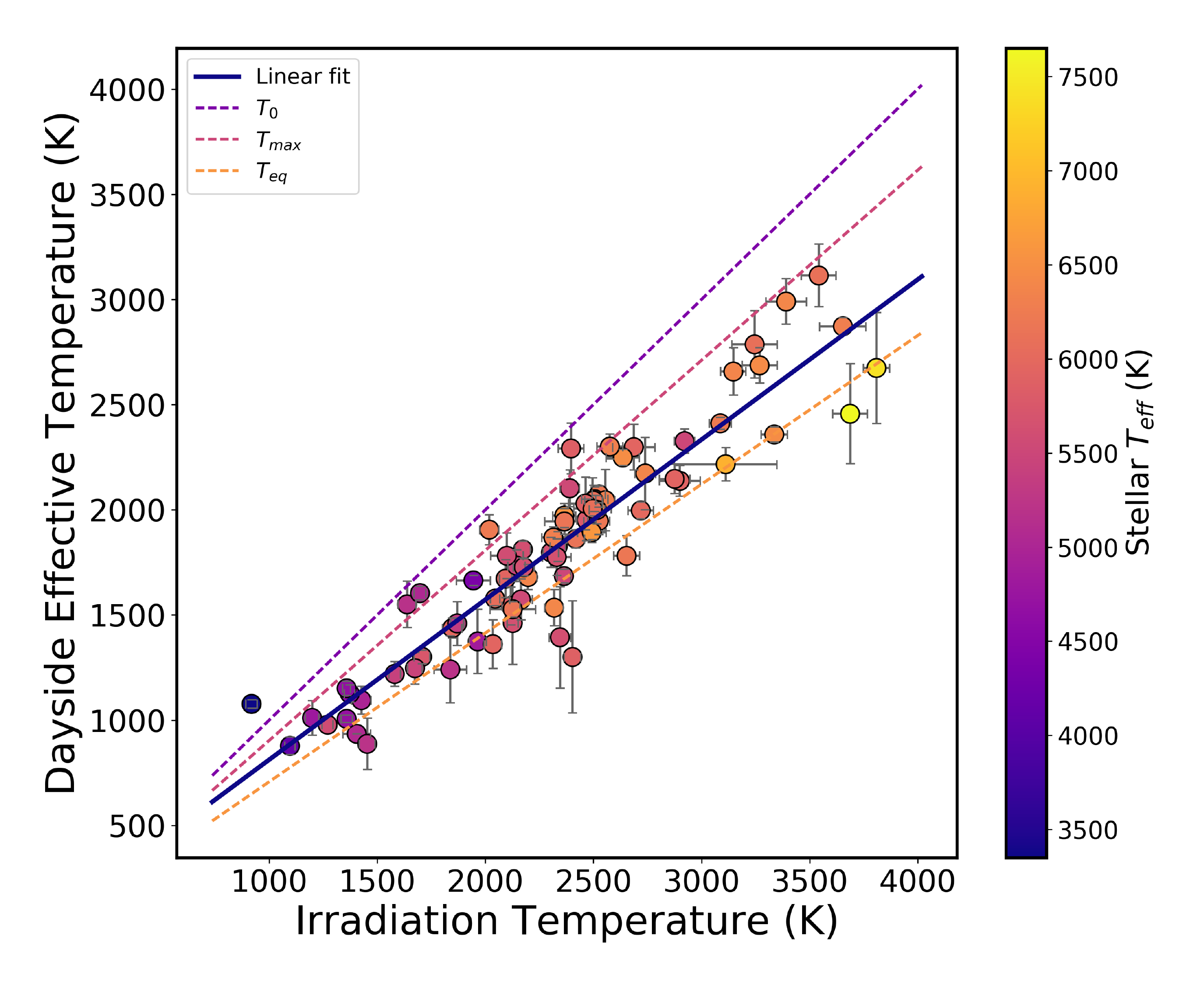}
    \caption{Dayside effective temperature ($<T_B>$) vs the theoretical irradiation temperature ($T_{eq}$) with zero albedo and full redistribution, similar to \citet{Schwartz2015}, but with 28 more planets. We also plot the expected irradiation temperature ($T_0$), the equilibrium temperature with zero albedo ($T_{eq}$), and the maximum dayside temperature ($T_{max}$). The color scale is the effective temperature of the host star in Kelvin.}
    \label{P2:fig:Td}
\end{figure}

Figure \ref{P2:fig:Td} presents the results for weighted mean effective temperature using the PHOENIX model calculations of the brightness temperatures. We find a slope of $0.76\pm0.05$, which is consistent with equilibrium temperature ($1\sigma$) and inconsistent with the $0.87\pm0.05$ of \citet{Schwartz2015}. However, with brightness temperatures calculated without integration over the bandpass and with blackbodies for the star we are able to retrieve a slope of $0.81\pm0.05,$ which is in statistical agreement (0.9$\sigma$) with their trend. 

Since we are able to retrieve the results using blackbodies, we conclude that the discrepancy is a result of careful use of stellar models and integration over the bandpasses and not of the differences in the sample sizes. More importantly, our findings do not support the findings presented in \citet{Schwartz2015} as we do not find that the effective temperature trend with irradiation temperature increases disproportionately. This means that we do not think the effective temperature calculated in this way tells us anything about the redistribution in the hottest planets. On the other hand, in Figure \ref{P2:fig:Tbs}, we find that the 4.5~$\mu$m brightness temperature is deviating from equilibrium, likely due to the strong CO opacity appearing in emission. This does support the hypothesis that these hottest planets are exhibiting different behaviors, but it is not expected to be captured in the effective temperatures since the weighted mean of the two brightness temperatures is likely muting this deviation.

\section{KELT-9b Eclipse: the hottest of the UHJs}
\label{P2:app:kelt9b}

A question arises of  whether  the trends presented  in Section \ref{P2:sec:disc} hold at even higher temperatures. To test this we include the 4.5~$\mu$m eclipse depth of the hottest of the UHJs, KELT-9b \citep{Gaudi2017}. Significantly hotter than any other ultra-hot Jupiter, KELT-9b is the hottest gas giant planet known. A 1.48-day orbital period around its A-type host star of 10170K makes it the most highly irradiated planet with an equilibrium temperature of 4050K \citep{Gaudi2017}. At these temperatures, the planet itself is similar to a K4 star;  its atmosphere is subject to molecular dissociation, leaving behind atomic metals such as iron and titanium \citep{Hoeijmakers2018, Hoeijmakers2019}. 

\begin{figure*}
    \centering
    \includegraphics[width=\linewidth]{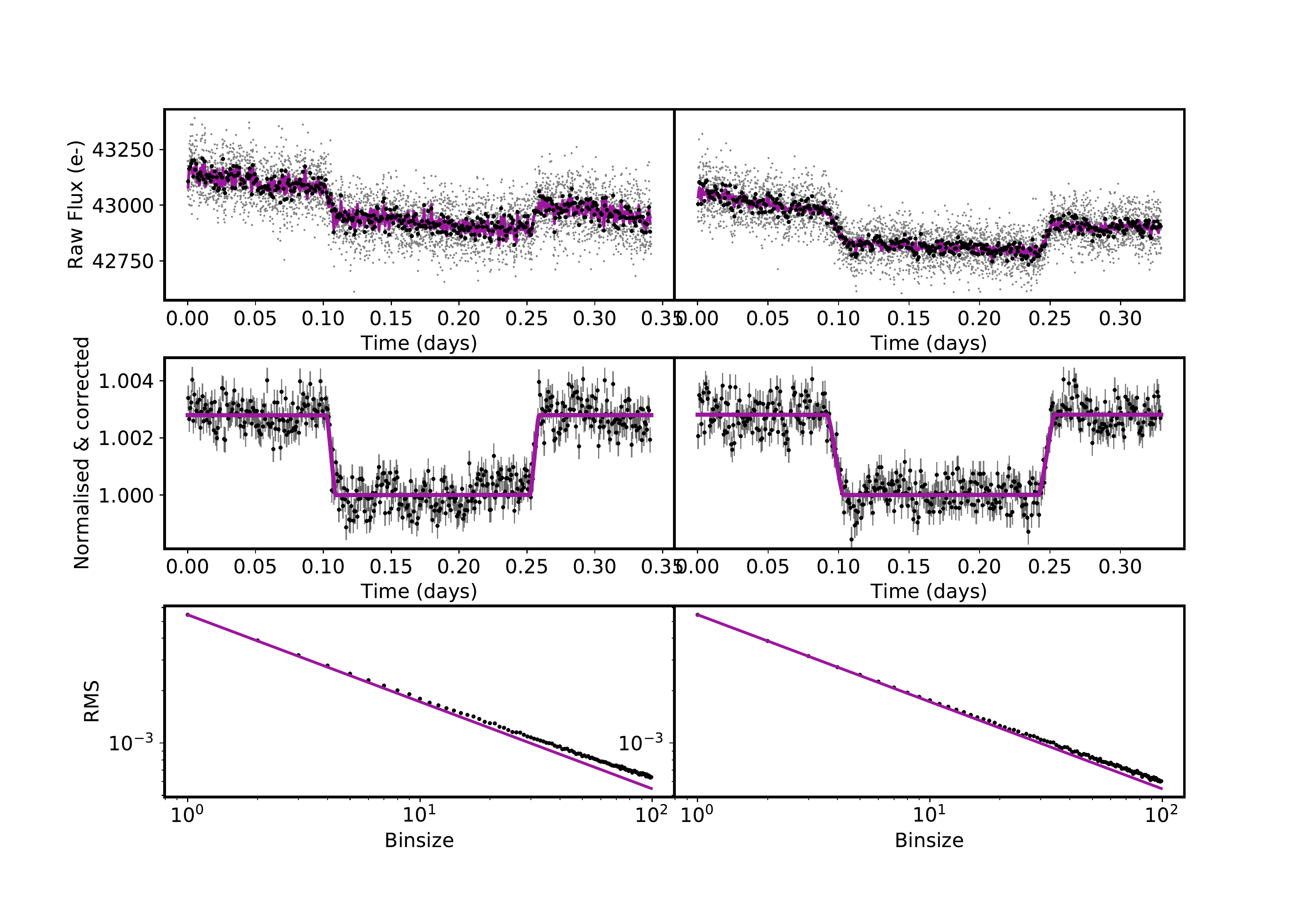}
    \caption{Eclipses of KELT-9b for AORs r67667712 (left panel) and r67667968 (right panel). The top row shows the raw photometric lightcurve with our best fit PLD model. The middle row shows the corrected lightcurves with the best fit eclipse model. The bottom row shows the RMS vs binsize of the data; since this closely follows the photon noise line ($\sqrt(N)$) we can see that we are capturing the systematics well.}
    \label{P2:fig:kelt9b}
\end{figure*}

We analyse two 4.5~$\mu$m eclipses of KELT-9b. The data were taken from the phase curve survey,   program ID 14059 lead by PI J Bean. We extracted the two secondary eclipses from the available phase curves. The analysis from raw data to eclipse depth values was done using our custom pipeline described in Baxter et al. (in prep.). In summary, we allow for different background correction methods, different centroiding methods, and different aperture radii to find the combination that gives the lowest $\chi^2$. We correct for the strong Spitzer systematics using Pixel Level Decorrelation \citep{Deming2015} and perform a full MCMC analysis using Batman \citep{Kreidberg2015} to fit for the eclipse parameters on the best photometric lightcurve. The raw photometry, the corrected lightcurves, and one of our statistical tests (RMS vs binsize, which characterizes how well we correct red noise) are  presented in Figure \ref{P2:fig:kelt9b}. The two eclipse depths (Fp/Fs) are calculated to be 2793 $\pm$ 44 (ppm) and 2809 $\pm$ 48 (ppm) for AORs r67667712 and r67667968, respectively. The eclipse depth used in the analysis is the mean of these two values 2801 $\pm$ 33 (ppm). This eclipse depth is used to calculate the brightness temperatures shown in Figure \ref{P2:fig:Tbs}. 

Our results disagree with the 4.5~$\mu$m eclipse depths presented in \citet{Mansfield2020} by 4.6$\sigma$. This significant difference is likely not due to any problems with the systematic correction algorithm, but is rather a result of the choice of baseline between eclipse and phase curve observations. Eclipse-only observations ignore phase variations, and can thus underestimate eclipse depths when the real phase variations are concave over the secondary eclipse \citep[e.g.,][]{Bell2019}. Since the large phase amplitude (0.601) presented in \citet{Mansfield2020} clearly demonstrates a concave phase variation around the ellipse, this is likely the cause of the discrepancy between the two sets of data analyses. However, since we do not see any trend with the phase curve offsets between the two Spitzer bandpasses (discussed in Section \ref{P2:sec:albedoredist}) we expect that any underestimation of the eclipse depth will apply to both 3.6~$\mu$m and 4.5~$\mu$m, and thus the deviation from the blackbody will be largely unaffected. Nevertheless, such an effect could be relevant for higher precision measurements with the James Webb Space Telescope.
 
Since we only have the available 4.5~$\mu$m measurement of KELT-9b, this data point is excluded from any of the fits in our analysis. Nevertheless, we can see that the brightness temperature deviates positively from equilibrium at 4.5~$\mu$m. However, like several of the hottest planets, the deviation is smaller than expected compared to the model predictions in Figure \ref{P2:fig:Tbs}. We interpret this as indicative of more complex physical processes happening in the atmosphere of this extreme object \citep{Bell2018, Komacek2018, Lothringer2018, Kitzmann2018, Mansfield2020}. For example, due to the high temperature on the dayside of KELT-9b, there could be less carbon monoxide available in the atmosphere due to it being dissociated \citep{Kitzmann2018}.
 
\end{appendix}

\clearpage
\onecolumn

\setlength{\tabcolsep}{3pt}

\begin{longtable}[h]{lllllllllll}

\caption{\label{P2:tab:longtab} Planetary eclipse depths from the literature, calculated equilibrium temperatures, calculated brightness temperatures, and deviations from blackbody using PHOENIX models and stellar parameters used to obtain the correct stellar models.}\\

\hline\hline
Planet & $(F_p/F_s)_{3.6}$ & $(F_p/F_s)_{4.5}$ &     $T_{eq_{a=0}}$ &           $T_{eff}$ &       logg  &           [Fe/H] & $T_{B_{3.6}}$ & $T_{B_{4.5}}$ &       devBB &            Ref.\\
 & (ppm) & (ppm) & (K) & (K) & (cgs) & (dex) & (K) & (K) & (\%) &             \\

\hline
\endfirsthead
\caption{continued.} \\
\hline\hline
Planet & $(F_p/F_s)_{3.6}$ & $(F_p/F_s)_{4.5}$ &     $T_{eq_{a=0}}$ &           $T_{eff}$ &       logg  &           [Fe/H] & $T_{B_{3.6}}$ & $T_{B_{4.5}}$ &       devBB &            Ref.\\
 & (ppm) & (ppm) & (K) & (K) & (cgs) & (dex) & (K) & (K) & (\%) &             \\
\hline
\endhead
\hline
\endfoot
HAT-P-32b     &    3640$\pm$160 &    4380$\pm$200 &   1785$\pm$32 &    6207$\pm$88 &  4.33$\pm$0.01 &  -0.04$\pm$0.08 &        2073$\pm$40 &        2023$\pm$46 &   0.006$\pm$0.026 &           1 \\
XO1b        &      860$\pm$70 &     1220$\pm$90 &   1207$\pm$30 &    5750$\pm$75 &   4.5$\pm$0.01 &   0.02$\pm$0.08 &        1301$\pm$32 &        1257$\pm$34 &  -0.001$\pm$0.012 &       2 \\
HAT-P-1b      &      800$\pm$80 &    1350$\pm$220 &   1306$\pm$33 &    5980$\pm$49 &  4.36$\pm$0.01 &   0.13$\pm$0.01 &        1437$\pm$47 &       1521$\pm$103 &   0.026$\pm$0.024 &        3 \\
WASP-39b     &     880$\pm$150 &     960$\pm$180 &   1118$\pm$35 &   5400$\pm$150 &    4.4$\pm$0.2 &   -0.12$\pm$0.1 &        1220$\pm$60 &        1066$\pm$63 &  -0.034$\pm$0.026 &         4 \\
HAT-P-18b     &     437$\pm$145 &     326$\pm$146 &    847$\pm$26 &    4803$\pm$80 &  4.57$\pm$0.04 &    0.1$\pm$0.08 &        1011$\pm$83 &         787$\pm$88 &   -0.04$\pm$0.025 &  5 \\
TrES2b      &    1270$\pm$210 &    2300$\pm$240 &   1498$\pm$32 &    5850$\pm$50 &  4.43$\pm$0.02 &   -0.15$\pm$0.1 &        1543$\pm$90 &        1712$\pm$81 &   0.063$\pm$0.034 &      6 \\
WASP-4b      &    3190$\pm$310 &    3430$\pm$270 &   1651$\pm$27 &    5436$\pm$34 &  4.46$\pm$0.05 &  -0.05$\pm$0.04 &        1825$\pm$72 &        1650$\pm$57 &  -0.049$\pm$0.042 &         7 \\
XO2b        &     810$\pm$170 &     980$\pm$200 &   1322$\pm$23 &    5340$\pm$32 &  4.48$\pm$0.05 &   0.45$\pm$0.02 &       1460$\pm$104 &       1346$\pm$104 &  -0.011$\pm$0.028 &       8 \\
WASP-1b      &    1170$\pm$160 &    2120$\pm$210 &   1876$\pm$69 &   6200$\pm$200 &    4.3$\pm$0.3 &     0.1$\pm$0.2 &        1781$\pm$95 &       2067$\pm$103 &   0.066$\pm$0.027 &       9 \\
HAT-P-26b     &        85$\pm$0 &      265$\pm$70 &    994$\pm$48 &    5079$\pm$88 &  4.56$\pm$0.06 &  -0.04$\pm$0.08 &          935$\pm$0 &        1067$\pm$90 &   0.011$\pm$0.007 &  5 \\
CoRoT-1 b   &    4150$\pm$420 &    4820$\pm$420 &   1900$\pm$81 &   5950$\pm$150 &   4.25$\pm$0.3 &   -0.3$\pm$0.25 &       2298$\pm$109 &       2236$\pm$102 &    0.006$\pm$0.06 &         10 \\
CoRoT-2 b   &    3550$\pm$200 &    5000$\pm$200 &   1537$\pm$40 &   5625$\pm$120 &  4.53$\pm$0.02 &   0.03$\pm$0.06 &        1811$\pm$40 &        1854$\pm$36 &   0.062$\pm$0.029 &         10 \\
HAT-P-17 b  &      118$\pm$39 &       149$\pm$... &    779$\pm$17 &    5246$\pm$80 &  4.53$\pm$0.02 &    0.0$\pm$0.08 &         807$\pm$54 &          704$\pm$... &    -0.009$\pm$... &  5 \\
HAT-P-19 b  &     620$\pm$140 &     620$\pm$140 &   1009$\pm$40 &   4990$\pm$130 &  4.54$\pm$0.05 &   0.23$\pm$0.08 &        1095$\pm$66 &         924$\pm$59 &  -0.036$\pm$0.023 &         4 \\
HAT-P-2 b   &      996$\pm$72 &     1031$\pm$61 &   1428$\pm$57 &    6290$\pm$60 &  4.16$\pm$0.08 &   0.14$\pm$0.08 &        2256$\pm$76 &        2065$\pm$62 &   -0.012$\pm$0.01 &          11 \\
HAT-P-20 b  &      615$\pm$82 &     1096$\pm$77 &    971$\pm$24 &    4595$\pm$80 &  4.63$\pm$0.02 &   0.35$\pm$0.08 &        1127$\pm$40 &        1131$\pm$26 &   0.014$\pm$0.013 &         12 \\
HAT-P-23 b  &    2480$\pm$190 &    3090$\pm$260 &   2051$\pm$71 &    5905$\pm$80 &  4.33$\pm$0.05 &   0.15$\pm$0.04 &        2137$\pm$73 &        2128$\pm$92 &   0.018$\pm$0.032 &       13 \\
HAT-P-3 b   &    1120$\pm$225 &     940$\pm$125 &   1158$\pm$34 &    5185$\pm$80 &  4.56$\pm$0.03 &   0.27$\pm$0.08 &       1550$\pm$110 &        1252$\pm$60 &  -0.053$\pm$0.028 &        14 \\
HAT-P-4 b   &    1420$\pm$160 &    1220$\pm$130 &   1694$\pm$47 &    5860$\pm$80 &  4.36$\pm$0.11 &   0.24$\pm$0.08 &       2291$\pm$120 &        1906$\pm$98 &  -0.041$\pm$0.021 &        14 \\
HAT-P-6 b   &     1170$\pm$80 &     1060$\pm$60 &   1673$\pm$42 &    6570$\pm$80 &  4.22$\pm$0.03 &  -0.13$\pm$0.08 &        1973$\pm$57 &        1681$\pm$43 &   -0.035$\pm$0.01 &        15 \\
HAT-P-7 b   &    1560$\pm$130 &    1900$\pm$110 &   2225$\pm$41 &    6389$\pm$17 &  47$\pm$0.06 &   0.26$\pm$0.08 &       2657$\pm$113 &        2704$\pm$92 &   0.016$\pm$0.017 &   16 \\
HAT-P-8 b   &     1310$\pm$85 &     1110$\pm$75 &   1772$\pm$48 &    6200$\pm$80 &  4.15$\pm$0.03 &   0.01$\pm$0.08 &        2050$\pm$58 &        1695$\pm$52 &  -0.045$\pm$0.012 &        15 \\
HD 149026 b &      400$\pm$30 &      340$\pm$60 &   1673$\pm$65 &    6160$\pm$50 &  4.28$\pm$0.05 &   0.36$\pm$0.08 &        1945$\pm$61 &       1603$\pm$122 &  -0.014$\pm$0.007 &      17 \\
HD 189733 b &    2560$\pm$140 &    2140$\pm$200 &   1200$\pm$22 &    5040$\pm$50 &  4.59$\pm$0.01 &  -0.03$\pm$0.08 &        1604$\pm$32 &        1298$\pm$45 &  -0.115$\pm$0.025 &    18 \\
HD 209458 b &     1190$\pm$70 &     1230$\pm$60 &   1446$\pm$19 &    6065$\pm$50 &  4.36$\pm$0.01 &    0.0$\pm$0.05 &        1577$\pm$33 &        1392$\pm$27 &   -0.033$\pm$0.01 &   19 \\
Kepler-12 b &    1370$\pm$200 &    1160$\pm$310 &   1481$\pm$31 &   5947$\pm$100 &  4.18$\pm$0.01 &   0.07$\pm$0.04 &        1672$\pm$91 &       1369$\pm$142 &  -0.058$\pm$0.038 &        20 \\
Kepler-17 b &    2500$\pm$300 &    3100$\pm$350 &   1745$\pm$39 &    5781$\pm$85 &  4.53$\pm$0.12 &    0.26$\pm$0.1 &        1952$\pm$98 &       1902$\pm$102 &   0.008$\pm$0.047 &        21 \\
Kepler-5 b  &    1030$\pm$170 &    1070$\pm$150 &   1807$\pm$35 &    6297$\pm$60 &   3.96$\pm$0.1 &   0.04$\pm$0.06 &       2045$\pm$146 &       1859$\pm$124 &  -0.016$\pm$0.023 &        22 \\
Kepler-6 b  &     690$\pm$270 &    1510$\pm$190 &   1504$\pm$21 &    5647$\pm$44 &  4.24$\pm$0.01 &   0.34$\pm$0.04 &       1462$\pm$196 &        1726$\pm$98 &   0.058$\pm$0.036 &         23 \\
KOI-13 b    &    1560$\pm$310 &    2220$\pm$230 &   2607$\pm$94 &   7650$\pm$250 &    4.2$\pm$0.5 &     0.2$\pm$0.2 &       2456$\pm$238 &       2716$\pm$164 &   0.044$\pm$0.039 &        24 \\
Qatar-1 b   &    1490$\pm$510 &    2730$\pm$490 &   1389$\pm$43 &   4861$\pm$125 &  4.54$\pm$0.02 &     0.2$\pm$0.1 &       1374$\pm$153 &       1470$\pm$108 &   0.068$\pm$0.077 &        25 \\
TrES-3 b    &    3460$\pm$350 &    3720$\pm$540 &   1629$\pm$32 &    5650$\pm$75 &  4.58$\pm$0.01 &  -0.19$\pm$0.08 &        1797$\pm$72 &       1624$\pm$103 &  -0.056$\pm$0.066 &        26 \\
TrES-4 b    &    1370$\pm$110 &    1480$\pm$160 &   1785$\pm$41 &    6200$\pm$75 &  4.06$\pm$0.02 &   0.14$\pm$0.09 &        1947$\pm$65 &        1793$\pm$90 &   -0.017$\pm$0.02 &        27 \\
WASP-10 b   &    1000$\pm$110 &    1460$\pm$160 &    960$\pm$24 &   4675$\pm$100 &    4.4$\pm$0.2 &    0.03$\pm$0.2 &        1151$\pm$35 &        1091$\pm$39 &  -0.007$\pm$0.021 &         11 \\
WASP-103 b  &    4458$\pm$383 &    5686$\pm$138 &   2505$\pm$78 &   6110$\pm$160 &  4.22$\pm$0.08 &   0.06$\pm$0.13 &       3114$\pm$149 &        3337$\pm$52 &    0.088$\pm$0.04 &      28 \\
WASP-12 b   &    4210$\pm$110 &    4280$\pm$120 &   2584$\pm$91 &   6300$\pm$150 &   4.38$\pm$0.1 &     0.3$\pm$0.1 &        2872$\pm$40 &        2649$\pm$43 &  -0.034$\pm$0.016 &      29 \\
WASP-121 b  &    3150$\pm$103 &    4510$\pm$107 &   2359$\pm$61 &   6459$\pm$140 &  4.24$\pm$0.01 &   0.13$\pm$0.09 &        2358$\pm$36 &        2591$\pm$35 &    0.09$\pm$0.015 &                 25 \\
WASP-14 b   &     1870$\pm$70 &    2240$\pm$180 &   1864$\pm$60 &   6475$\pm$100 &   4.07$\pm$0.2 &     0.0$\pm$0.2 &        2248$\pm$39 &        2221$\pm$93 &   0.007$\pm$0.019 &         30 \\
WASP-18 b   &    3000$\pm$200 &    3900$\pm$200 &   2398$\pm$73 &   6400$\pm$100 &  4.37$\pm$0.04 &    0.0$\pm$0.09 &       2990$\pm$109 &       3231$\pm$104 &   0.063$\pm$0.028 &        31 \\
WASP-19 b   &    4830$\pm$250 &    5720$\pm$300 &   2066$\pm$46 &   5500$\pm$100 &    4.5$\pm$0.2 &   0.02$\pm$0.09 &        2326$\pm$57 &        2270$\pm$63 &    0.02$\pm$0.039 &       32 \\
WASP-2 b    &     830$\pm$350 &    1690$\pm$170 &   1300$\pm$71 &   5200$\pm$200 &  4.54$\pm$0.04 &     0.1$\pm$0.2 &       1241$\pm$158 &        1350$\pm$51 &   0.048$\pm$0.046 &       9 \\
WASP-24 b   &    1590$\pm$130 &    2020$\pm$180 &   1769$\pm$39 &   6075$\pm$100 &  4.26$\pm$0.17 &     0.0$\pm$0.1 &        2044$\pm$73 &        2044$\pm$92 &   0.013$\pm$0.022 &          33 \\
WASP-33 b   &    2600$\pm$500 &    4100$\pm$200 &   2694$\pm$53 &   7430$\pm$100 &    4.3$\pm$0.2 &     0.1$\pm$0.2 &       2674$\pm$264 &        3202$\pm$98 &   0.119$\pm$054 &         34 \\
WASP-43 b   &    3460$\pm$130 &    3820$\pm$150 &   1375$\pm$79 &   4400$\pm$200 &  4.65$\pm$0.05 &  -0.05$\pm$0.17 &        1664$\pm$24 &        1497$\pm$24 &   -0.053$\pm$0.02 &         35 \\
WASP-48 b   &    1760$\pm$130 &    2140$\pm$200 &   2033$\pm$68 &   5920$\pm$150 &  4.03$\pm$0.03 &  -0.12$\pm$0.12 &        2147$\pm$70 &       2113$\pm$101 &   0.008$\pm$0.024 &       13 \\
WASP-5 b    &    1970$\pm$280 &    2370$\pm$240 &   1742$\pm$68 &   5880$\pm$150 &   4.4$\pm$0.04 &   09$\pm$0.09 &       2030$\pm$125 &        1969$\pm$98 &   0.002$\pm$0.038 &         36 \\
WASP-6 b    &     940$\pm$190 &    1150$\pm$220 &   1184$\pm$32 &   5450$\pm$100 &    4.6$\pm$0.2 &   -0.2$\pm$0.09 &        1247$\pm$75 &        1134$\pm$72 &  -0.022$\pm$0.032 &         4 \\
WASP-67 b   &     220$\pm$130 &     800$\pm$180 &   1028$\pm$32 &   5200$\pm$100 &  4.35$\pm$0.15 &  -07$\pm$0.09 &        887$\pm$122 &        1042$\pm$73 &   0.039$\pm$0.027 &         4 \\
WASP-69 b   &      421$\pm$29 &      463$\pm$39 &    961$\pm$21 &    4700$\pm$50 &   4.5$\pm$0.15 &   0.15$\pm$0.08 &        1006$\pm$17 &         864$\pm$19 &  -0.024$\pm$0.006 &  5 \\
WASP-8 b    &    1130$\pm$180 &      690$\pm$70 &    927$\pm$27 &    5600$\pm$80 &    4.5$\pm$0.1 &   0.17$\pm$0.07 &        1573$\pm$90 &        1144$\pm$39 &  -0.078$\pm$0.021 &       37 \\
WASP-80 b   &     455$\pm$100 &      944$\pm$65 &    775$\pm$25 &   4150$\pm$100 &    4.6$\pm$0.2 &  -0.14$\pm$0.16 &         878$\pm$42 &         875$\pm$16 &    0.01$\pm$0.016 &         38 \\
XO-3 b      &     1010$\pm$40 &     1580$\pm$36 &   2046$\pm$40 &    6429$\pm$50 &  3.95$\pm$0.06 &  -0.18$\pm$0.03 &        1814$\pm$29 &        1972$\pm$21 &   0.033$\pm$0.006 &       39 \\
XO-4 b      &      560$\pm$90 &     1350$\pm$85 &   1639$\pm$35 &    6397$\pm$70 &  4.18$\pm$0.07 &  -0.04$\pm$0.03 &        1535$\pm$86 &        1957$\pm$60 &   0.061$\pm$0.013 &        15 \\
HAT-P-13 b  &     851$\pm$107 &    1090$\pm$124 &   1648$\pm$53 &    5653$\pm$90 &  4.13$\pm$0.04 &   0.41$\pm$0.08 &        1775$\pm$87 &        1728$\pm$89 &   0.003$\pm$0.017 &                25 \\
HAT-P-30 b  &    1584$\pm$107 &    1825$\pm$147 &   1637$\pm$43 &    6304$\pm$88 &   4.36$\pm$0.3 &   0.13$\pm$0.08 &        1868$\pm$51 &        1763$\pm$65 &  -0.012$\pm$0.019 &                25 \\
HAT-P-33 b  &    1603$\pm$127 &    1835$\pm$199 &   1780$\pm$34 &    6446$\pm$88 &  4.15$\pm$0.01 &   0.05$\pm$0.08 &        2000$\pm$67 &        1901$\pm$98 &  -0.009$\pm$0.024 &                25 \\
HAT-P-40 b  &     988$\pm$168 &    1057$\pm$145 &   1765$\pm$66 &   6080$\pm$100 &  3.93$\pm$0.02 &    0.22$\pm$... &       2005$\pm$146 &       1840$\pm$119 &  -0.012$\pm$0.023 &                25 \\
HAT-P-41 b  &    1829$\pm$319 &    2278$\pm$177 &   1937$\pm$44 &   6390$\pm$100 &  4.14$\pm$0.02 &    0.21$\pm$0.1 &       2173$\pm$172 &        2179$\pm$88 &   0.014$\pm$0.037 &                25 \\
KELT-2 A b  &      650$\pm$38 &      678$\pm$47 &   1710$\pm$31 &    6151$\pm$50 &  4.03$\pm$0.02 &  -0.02$\pm$0.07 &        1862$\pm$44 &        1679$\pm$52 &  -0.012$\pm$0.006 &                25 \\
KELT-3 b    &     1766$\pm$97 &    1656$\pm$104 &   1822$\pm$44 &    6304$\pm$49 &   4.2$\pm$0.03 &   0.05$\pm$0.08 &        2300$\pm$59 &        2006$\pm$62 &  -0.038$\pm$0.014 &                25 \\
WASP-100 b  &     1267$\pm$98 &    1720$\pm$119 &  2200$\pm$171 &   6900$\pm$120 &  4.35$\pm$0.17 &   -0.03$\pm$0.1 &        2216$\pm$79 &        2337$\pm$88 &   0.024$\pm$0.016 &                25 \\
WASP-101 b  &    1161$\pm$111 &    1194$\pm$113 &   1554$\pm$40 &   6380$\pm$120 &  4.31$\pm$0.08 &    0.2$\pm$0.12 &        1680$\pm$61 &        1492$\pm$58 &  -0.029$\pm$0.017 &                25 \\
WASP-104 b  &    1709$\pm$195 &    2643$\pm$303 &   1516$\pm$43 &   5475$\pm$127 &   4.5$\pm$0.02 &   0.32$\pm$0.09 &        1734$\pm$76 &        1828$\pm$98 &    0.05$\pm$0.037 &                25 \\
WASP-131 b  &      364$\pm$97 &      282$\pm$78 &   1439$\pm$36 &   5950$\pm$100 &    3.9$\pm$0.1 &  -0.18$\pm$0.08 &       1361$\pm$115 &        1077$\pm$96 &  -0.023$\pm$0.014 &                25 \\
WASP-36 b   &     913$\pm$578 &    1948$\pm$544 &   1699$\pm$45 &   5881$\pm$136 &   4.5$\pm$0.01 &  -0.31$\pm$0.12 &       1300$\pm$267 &       1475$\pm$168 &   0.064$\pm$0.089 &                25 \\
WASP-46 b   &    1360$\pm$701 &    4446$\pm$589 &   1658$\pm$55 &   5620$\pm$160 &  4.49$\pm$0.02 &  -0.37$\pm$0.13 &       1394$\pm$241 &       1968$\pm$129 &   0.258$\pm$0.101 &                25 \\
WASP-62 b   &    1616$\pm$146 &    1359$\pm$130 &   1427$\pm$35 &    6230$\pm$80 &   4.45$\pm$0.1 &   0.04$\pm$0.06 &        1906$\pm$71 &        1568$\pm$63 &   -0.061$\pm$0.02 &                25 \\
WASP-63 b   &      552$\pm$95 &     533$\pm$128 &   1531$\pm$45 &   5550$\pm$100 &  4.01$\pm$0.03 &   0.08$\pm$0.07 &        1573$\pm$97 &       1347$\pm$123 &  -0.018$\pm$0.017 &                25 \\
WASP-64 b   &    2859$\pm$270 &    2071$\pm$471 &   1690$\pm$52 &   5550$\pm$150 &  4.39$\pm$02 &  -08$\pm$0.11 &        2102$\pm$87 &       1610$\pm$159 &  -0.129$\pm$0.055 &                25 \\
WASP-65 b   &    1587$\pm$245 &     724$\pm$318 &   1485$\pm$59 &   5600$\pm$100 &   4.25$\pm$0.1 &  -07$\pm$0.07 &       1781$\pm$108 &       1160$\pm$177 &  -0.125$\pm$0.041 &                25 \\
WASP-74 b   &     1446$\pm$66 &    2075$\pm$100 &   1923$\pm$53 &   5990$\pm$110 &  4.39$\pm$0.07 &   0.39$\pm$0.13 &        1997$\pm$39 &        2106$\pm$51 &   0.034$\pm$0.012 &                25 \\
WASP-76 b   &     2645$\pm$63 &     3345$\pm$82 &   2183$\pm$47 &   6250$\pm$100 &    4.4$\pm$0.1 &    0.23$\pm$0.1 &        2411$\pm$28 &        2471$\pm$33 &    0.034$\pm$0.01 &                25 \\
WASP-77 A b &     1845$\pm$94 &    2362$\pm$127 &   1671$\pm$31 &    5500$\pm$80 &  4.33$\pm$0.08 &    0.0$\pm$0.11 &        1685$\pm$32 &        1628$\pm$37 &   0.002$\pm$0.016 &                25 \\
WASP-78 b   &    2001$\pm$218 &    2013$\pm$351 &   2295$\pm$88 &   6100$\pm$150 &    4.1$\pm$0.2 &  -0.35$\pm$0.14 &       2787$\pm$160 &       2565$\pm$255 &  -0.019$\pm$0.041 &                25 \\
WASP-79 b   &     1394$\pm$88 &    1783$\pm$106 &   1762$\pm$53 &   6600$\pm$100 &   4.2$\pm$0.15 &    0.03$\pm$0.1 &        1893$\pm$49 &        1882$\pm$54 &   0.008$\pm$0.014 &                25 \\
WASP-87 b   &    2077$\pm$127 &    2705$\pm$137 &   2311$\pm$68 &   6450$\pm$120 &  4.32$\pm$0.21 &   -0.41$\pm$0.1 &        2687$\pm$85 &        2863$\pm$87 &    0.04$\pm$0.019 &                25 \\
WASP-94 A b &      867$\pm$59 &      995$\pm$93 &   1504$\pm$77 &    6170$\pm$80 &  4.27$\pm$0.07 &   0.26$\pm$0.15 &        1527$\pm$36 &        1398$\pm$50 &  -0.016$\pm$0.011 &                25 \\
WASP-97 b   &     1359$\pm$84 &    1534$\pm$101 &   1540$\pm$42 &   5640$\pm$100 &  4.45$\pm$0.08 &   0.23$\pm$0.11 &        1727$\pm$40 &        1590$\pm$44 &  -0.017$\pm$0.014 &                25 \\
KELT-9 b    &         ...$\pm$... &     2802$\pm$33 &  4051$\pm$199 &  10170$\pm$450 &  4.09$\pm$0.01 &   -0.03$\pm$0.2 &            ...$\pm$... &            ...$\pm$... &       ...$\pm$... &     40 \\
\end{longtable}
\tablebib{
(1) \citet{Zhao2014}; (2) \citet{Machalek2008}; (3) \citet{Todorov2010}; (4) \citet{Kammer2015};
(5) \citet{Wallack2018inprep}; (6) \citet{ODonovan2010}; (7) \citet{Beerer2011}; (8) \citet{Machalek2009};
(9) \citet{Wheatley2010}; (10) \citet{Deming2011}; (11) \citet{Lewis2013}; (12) \citet{Deming2015};
(13) \citet{ORourke2014}; (14) \citet{Todorov2013}; (15) \citet{Todorov2012}; (16) \citet{Christiansen2010};
(17) \citet{Stevenson2012}; (18) \citet{Charbonneau2008}; (19) \citet{Diamond-Lowe2014}; (20) \citet{Fortney2011}; 
(21) \citet{Desert2011c}; (22) \citet{Desert2011b}; (23) \citet{Desert2011}; (24) \citet{Shporer2014};
(25) \citet{Garhart2020}; (26) \citet{Fressin2010}; (27) \citet{Knutson2009b}; (28) \citet{Kreidberg2018}; 
(29) \citet{Stevenson2014}; (30) \citet{Blecic2013}; (31) \citet{Nymeyer2011}; (32) \citet{Anderson2013};
(33) \citet{Smith2012}; (34) \citet{Deming2012}; (35) \citet{Blecic2014}; (36) \citet{Baskin2013};
(37) \citet{Cubillos2013}; (38) \citet{Triaud2015}; (39) \citet{Machalek2010}; (40) This work.;
}

\end{document}